\documentclass[a4paper,11pt]{article}
\usepackage{jhepmod} 
\usepackage{lineno,xcolor,ulem}

\definecolor{green3}{RGB}{44,160,44}

\usepackage{booktabs}  
\usepackage{caption}   
\usepackage{siunitx}
\usepackage{booktabs}
\usepackage[table]{xcolor}
\usepackage{hhline}
\usepackage{makecell}

\title{Real-time Gravitational Wave Response in Thermal Spinning fields}

\author[1]{Atsuhisa Ota,}
\author[2]{Hui-Yu Zhu,}
\author[2]{and Yuhang Zhu}
\affiliation[1]{Department of Physics and Chongqing Key Laboratory for Strongly Coupled Physics, \\
Chongqing University, Chongqing 401331, People's Republic of China}
\affiliation[2]{Cosmology, Gravity and Astroparticle Physics Group,\\
        Center for Theoretical Physics of the Universe,\\
        Institute for Basic Science, Daejeon 34126, Korea}
\emailAdd{aota@cqu.edu.cn}
\emailAdd{hzhuav@ibs.re.kr}
\emailAdd{yhzhu@ibs.re.kr}

\abstract{
We study how the spin content of the thermal plasmas affects the propagation of gravitational waves in a radiation-dominated universe. As a simple but representative setup, we consider conformal scalar, Weyl fermion, and Maxwell fields that provide the background radiation, and we ask whether the resulting damping and phase shift of gravitational waves retain any memory of their spins.
We revisit this question in a real-time quantum-field-theoretic framework, where the stress tensor splits into a background part, a dynamical (history-dependent) response, and local contact terms, with an additional on-shell projection fixed by the Friedmann equation. 
We find that the dynamical spin-dependent response arises on a short time scale characterized by the radiation temperature, which is exactly canceled by the local responses.
As a result, the remaining long-time response is universal and consistent with kinetic theory in the hard thermal limit. Although the underlying mechanism exhibits strong spin dependence, it leaves no observable imprint on the large-scale effective dynamics of gravitational waves in this setup. 
}

\begin{document}
\maketitle

\section{Introduction}

It has long been known that a thermal medium can influence the propagation of quantum fields~\cite{Bellac:2011kqa,Laine:2016hma}.
A particularly well-developed example is the hard thermal loop (HTL) theory of gauge fields, which provides an effective description of how the thermal medium modifies long-wavelength field propagation~\cite{Braaten:1989mz,Frenkel:1989br,Braaten:1990it,Braaten:1991gm,Taylor:1990ia}. 
In the soft limit of the external gauge field, the HTL self-energy obtained in thermal quantum field theory agrees with the prediction of classical linear response based on collisionless kinetic theory~\cite{Blaizot:1993zk,Kelly:1994ig,Blaizot:2001nr}.
Gravitational analogues have also been studied in the thermal Minkowski limit~\cite{Rebhan:1990yr,Brandt:1991qn,Brandt:1992dk,Brandt:1993bj,Nachbagauer:1994jy}. 
This line of work has established the structure of the thermal graviton HTL in the Euclidean framework, but with a few exceptions, it has been restricted to minimally coupled scalar fields~\cite{deAlmeida:1993wy}.

The kinetic counterpart of the gravitational HTL is particularly well suited to cosmology. 
In this approach, developed by Weinberg~\cite{Weinberg:2003ur} and by Rebhan and Schwarz~\cite{Rebhan:1994zw}, the effective dynamics of long-wavelength gravitational waves propagating in a thermal medium can be obtained by solving a nonlocal integro-differential equation. 
The anisotropic stress is history-dependent and proportional to the number of on-shell degrees of freedom. 
By using the background Friedmann equation, the explicit dependence on particle spin can be removed, so that the induced stress takes a universal form in an evolving background.

More recently, real-time analyses based on the in-in formalism~\cite{Weinberg:2005vy,Adshead:2009cb} have been applied to gravitons propagating on a time-dependent, radiation-dominated background~\cite{Ota:2023iyh,Frob:2025sfq,Ota:2024idm,Ota:2025yeu}. 
In this framework, one directly computes the gravitational-wave power spectrum in the operator formalism, and finds the existence of tachyonic mass terms as well as secular growth. 
However, symmetry arguments show that such pathologies can be removed once the constraints of general covariance are implemented correctly~\cite{Ota:2025rll}. As emphasized in this line of work, the effects of thermalized spinning particles on the propagation of gravitational waves are reduced to computing the thermal expectation value of unequal-time stress tensor commutation relation.

As has been widely noticed in the literature on Weyl anomalies, the expectation values of the stress tensors depend strongly on the spin content of the underlying quantum fields \cite{Christensen:1977jc, Brown:1976wc, Brown:1977pq,Capper:1974ic,Duff:1977ay, Duff:2000mt,Osborn:1993cr,Duff:1993wm,Buchel:2009sk}. In conformal field theory~(CFT), for instance, the two-point function of the stress tensor is almost completely fixed by symmetry. Its dynamical tensorial structure is universal and does not depend on the spin of the fields, while its overall coefficient, usually denoted $C_T$, is a spin-dependent constant. 
Although the zero-temperature result cannot simply be extrapolated to finite temperature, it suggests that the dependence on spin resides entirely in an overall coefficient of the response and not in the detailed dynamical structures.
Indeed, explicit calculations show that stress tensor correlators for thermal spinning fields contain similar spin structures in a short-time response.
This CFT argument is, in practice, relevant to the gravitational-wave response in a radiation-dominated universe.
One might therefore expect that, even at finite temperature, the linear response of the stress tensor would inherit the same dependence on spin.
This appears to be at odds with the kinetic descriptions, which predict a universal, spin-independent nonlocal response once the background Friedmann equations are used. 
Is this a failure of kinetic theory, or does thermal quantum field theory itself reproduce the kinetic prediction in a nontrivial way?

In this paper we address this question in a concrete setting. 
We consider thermal field theories of spin~0, spin~1/2, and spin~1 fields, and compute the real-time linear response of the stress tensor to a gravitational-wave perturbation. 
Working directly in the in-in formalism, we derive the graviton self-energy in the dynamical background driven by the thermal fields themselves. 
We show that the nonlocal part of the response, which governs the history-dependent damping and phase shift of primordial gravitational waves, agrees with the universal result of kinetic theory and is independent of spin. 
This is because the expected spin dependence associated with $C_T$, which resides in the short-time response even at finite temperature, is exactly canceled by other local contributions.

\section{Basic setup and summary of the results}

In this work, we examine the effective dynamics of gravitational waves in a radiation-dominated universe including thermal self-energy. In particular, we are interested in cases where thermal fields with spin account for the thermal radiation and propagate in hard loops of the graviton self-energy in the expanding universe. In this section, we provide a brief introduction to the setup and a quick summary of the results.

\subsection{Setup}
We write the perturbed FLRW metric as
\begin{align}
	g^{\mu\nu} = a^{-2} (\eta^{\mu\nu} + \kappa h^{\mu\nu}) ~ .\label{defmetric}
\end{align}
with the gravitational coupling $\kappa=\sqrt{32\pi G}=2/M_{\rm{pl}}$. $h^{\mu\nu}$ is regarded as a tensor field in the background spacetime, and its indices are raised and lowered using the flat spacetime metric.
We use this convention because the Jacobian for the change of variables $g^{\mu\nu}\to h^{\mu\nu}$ is trivial: 
\begin{align}
	\frac{\delta}{\delta g^{\mu\nu}} \propto \frac{\delta}{\delta h^{\mu\nu}} ~ .
\end{align}
Consider a conformal scalar field~(spin 0), a Weyl fermion~(spin 1/2), and a Maxwell field~(spin 1) that sustain a radiation-dominated universe, while gravity is described by general relativity. We expand the action in powers of $\kappa$ and use interaction-picture perturbation theory for the gravitational wave dynamics. For simplicity, we assume a collisionless plasma, meaning that we consider only interactions between the quantum fields and the gravitational waves. Interactions among the quantum fields themselves are ignored, and thermal states are introduced through an initial thermal ensemble. 
In this sense, our setup is not hydrodynamical and follows the typical setup in HTL analyses.
In practice, our setting is well suited to describing relic radiation after decoupling, such as cosmic neutrinos.

The effective equation of motion for $h_{\mu\nu}= \eta_{\mu \alpha}\eta_{\nu\beta}h^{\alpha \beta}$ is obtained by linearizing the Einstein equations
\begin{align}
	G_{\mu\nu} = \frac{\kappa^2}{4}T_{\mu\nu},\label{Einsteineq}
\end{align}
where $G_{\mu\nu}$ is the Einstein tensor and $T_{\mu\nu}$ is the effective stress tensor obtained after integrating out UV degrees of freedom, which in the present case are the thermal conformal fields. At this order, only the linear response (i.e. $\mathcal{O}(\kappa)$) of the gravitational perturbations to the effective stress tensor is relevant. 

\subsection{Kinetic theory}
Weinberg analyzed the same issue in classical kinetic theory~\cite{Weinberg:2003sw} (see also \cite{Rebhan:1994zw,Francisco:2016rtf}). For a given phase-space distribution function $F$, the stress tensor can be written as
\begin{align}
	T_{\mu\nu} = \int \frac{d^4 p}{(2\pi )^4\sqrt{-g}}  p_\mu p_\nu F.\label{kinth}
\end{align}
The linear gravitational response of $F$ is then obtained by solving the Vlasov equation
\begin{align}
	\frac{dF}{d\lambda} = 0,
\end{align}
with the affine parameter $\lambda$.
This equation can be integrated formally, order by order in $\kappa$. 
The detailed structure of the kinetic graviton HTL is discussed in appendix~\ref{sec:Kinth}.
Let us denote the relativistic number of degrees of freedom of the spin-$s$ fields as $g_s$.
The effective relativistic number of degrees of freedom is then
\begin{align}
	g_* = \sum_{s=0,1,\cdots} g_s + \frac{7}{8}\sum_{s=\frac{1}{2},\cdots} g_s.
\end{align} 
Since the kinematics is insensitive to spin, the effective stress tensor is determined by the total phase-space distribution function
\begin{align}
	F = g_* F_0,\label{fisum}
\end{align}
where $F_0$ is the spin-0 phase-space distribution.
In this framework, the linear response in \eqref{kinth} is proportional to the number of on-shell channels, $g_*$, and to the fourth power of the temperature~\footnote{See Eq.~\eqref{kernelexpand} for the more explicit dependence on the energy density.}.

One might think that the effective dynamics is sensitive to $g_*$. However, the radiation temperature $T$ is not a free parameter in the expanding universe.
We must solve the background Friedmann equation simultaneously:
\begin{align}
	3H^2 = \frac{\kappa^2}{4} \rho.
\end{align}
One can express the energy density in terms of the Hubble parameter, so that the dependence on $g_*$ is implicit after matching the radiation temperature via the Friedmann equation.
Thus, kinetic theory predicts a universal anisotropic stress, independent of the spin content of the radiation.

\subsection{QFT-based approach}
The same problem can be addressed in quantum field theory. Let $\hat T_{\mu\nu}$ be the stress tensor operator, $\hat T_{\mu\nu}^I \equiv \lim\limits_{\kappa \to 0}\hat T_{\mu\nu}$ its interaction-picture counterpart, $\hat H_{\rm int}$ the leading interaction Hamiltonian, and $\langle \cdots \rangle$ the average over the initial ensemble of the quantum fields. In the linear response, or in the in-in formalism of quantum field theory, the effective stress tensor is expressed as 
\begin{align}
	T_{\mu\nu} = T^{\rm bg}_{\mu\nu}  + T^{\rm dyn.}_{\mu\nu} + T^{\rm cont.}_{\mu\nu}+ \mathcal O(\kappa^2),\label{deflinres}
\end{align}
where we defined
\begin{align}
	T^{\rm bg}_{\mu\nu} &=\langle \hat T^{I}_{\mu\nu} \rangle,\label{def:bgT}
	\\
	T^{\rm dyn.}_{\mu\nu} &= i \int^\tau_{\tau_0} d\tau_1 \left \langle \left[ \hat H_{\rm int}(\tau_1), \hat T^{I}_{\mu\nu} \right] \right\rangle,\label{def:dyT}
	\\
	T^{\rm cont.}_{\mu\nu} &=  \int d^4 x \left \langle \frac{\delta \hat T_{\mu\nu}}{\delta g^{\rho\sigma}(x)} \right \rangle_{\kappa=0} \delta g^{\rho \sigma}(x).\label{def:ctT}
\end{align}
$\langle \cdots\rangle$ is the thermal average defined in a background spacetime.
The ensemble must be the global Gibbs ensemble. If one instead takes a local ensemble perturbed by the metric, the Ward identities for Weyl and diffeomorphism symmetries are violated~\cite{Ota:2025rll}. 
$T^{\rm bg}_{\mu\nu}$ is the leading free evolution of the stress tensor unperturbed by the metric. $T^{\rm dyn.}_{\mu\nu}$ gives the dynamical evolution generated by the leading interaction, and it represents a variation through canonical variables.
This term is dynamical and nonlocal. 
$T^{\rm cont.}_{\mu\nu}$ is the contact term, arising from the explicit metric dependence of the stress tensor.
The contact response is non-dynamical and local.

Traditional analyses of graviton HTLs have focused mainly on the slow part of the dynamical response in $T^{\rm dyn.}_{\mu\nu}$, typically within the imaginary time formalism. To our knowledge, local terms have often been discarded through \textit{ad hoc} prescriptions~\cite{Rebhan:1990yr}.

The HTL self-energy is essentially classical
%
~\cite{Blaizot:1993zk,Kelly:1994ig,Blaizot:2001nr}. 
For gravitons, the same equivalence has been demonstrated for real scalar fields~\cite{Francisco:2016rtf}.
For spinning hard particles, it has been suggested that one can simply weight by the relativistic degrees of freedom, consistent with Eq.~\eqref{fisum}~\cite{Rebhan:1990yr}.
In this work, we take a different perspective.
The interaction Hamiltonian is generally expressed as
\begin{align}
	H_{\rm int} = \frac{\kappa}{2} \int d^3 x a^2 h^{\mu\nu}\hat T^I_{\mu\nu}.\label{defHint}
\end{align}
We adopt the sign convention of Eq.~\eqref{defmetric} and
Eq.~\eqref{defHint} is consistent with 
\begin{align}
	\hat T_{\mu\nu} =-\frac{2}{\sqrt{-g}} \frac{\delta S}{\delta g^{\mu\nu}}.
\end{align}
The dynamical response term in Eq.~\eqref{deflinres} is then
\begin{align}
	T^{\rm dyn.}_{\mu\nu} = \frac{i\kappa}{2} \int^\tau_{\tau_0} d\tau_1 \int d^3x_1  a^2(\tau_1) \left \langle \left[  \hat T^{I}_{\rho \sigma}(\tau_1, \mathbf{x}_1), \hat T^{I}_{\mu\nu}(\tau, \mathbf{x}) \right] \right\rangle h^{\mu\nu}(\tau_1, \mathbf{x}_1).\label{TTcomut}
\end{align}
Thus, the task reduces to evaluating the ensemble average of an unequal-time commutator of radiation stress tensors.

\subsection{Spin-dependent correlation functions}
The stress tensor correlation functions appearing in Eq.~\eqref{TTcomut} can depend strongly on the spin content of the theory. As a familiar example, in a $d$-dimensional conformal field theory at zero temperature one finds~\cite{Duff:2004wh, Osborn:1993cr,Buchel:2009sk,OsbornCFTNotes}
\begin{align}
	\langle \hat T^{I}_{\mu\nu}(x) \hat T^{I}_{\rho\sigma}(0) \rangle  = \frac{C_T}{x^{2d}} I_{\mu\nu\rho\sigma},\label{cft}
\end{align}
with a traceless tensor $I_{\mu\nu\rho\sigma}$.
The vacuum expectation value of the stress tensor commutator also has the same overall factor $C_T$, which depends on spin.
For a spin-$s$ field in $d=4$ with multiplicity $n_s$, the well-known expression is
\begin{align}
	C_T = n_0 + 3 n_{\frac{1}{2}} + 12 n_{1}.\label{CTdef}
\end{align} 
This ratio $1:3:12$ is fixed by the spin algebra and is therefore independent of the internal structure of the correlation functions.
Eq.~\eqref{CTdef} suggests that the weight in $T^{\rm dyn.}_{\mu\nu}$ is $C_T$ from a field-theoretic point of view.
Although this result is derived at zero temperature, the underlying algebraic structure is expected to persist in the thermal case.
If the net effect depends on the spin content of the hard particles, then we expect a result different from the kinetic-theory case. To further clarify the possible discrepancy, we investigate the weight factor in a QFT approach in the real-time formalism.

\subsection{Key results}
In short, when~\eqref{Einsteineq} is linearized in a dynamical background, one finds
\begin{align}
	\widehat{\rm EoM}\,h_{\mu\nu} = \frac{\kappa^2}{4}T^{\rm lin.}_{\mu\nu}[h],\label{eomschem}
\end{align}
with the linear response of gravitational perturbations $T^{\rm lin.}_{\mu\nu}$ in the real-time formalism.
The precise identification of $T^{\rm lin.}_{\mu\nu}$ matters.
We find that $T^{\rm lin.}_{\mu\nu}$ is universal and coincides with the prediction of classical kinetic theory in the HTL limit.
However, the cancellation mechanism in quantum field theory is highly nontrivial: the spin-dependent coefficients, similar to those appearing in Eq.~\eqref{CTdef}, do arise at intermediate stages of the calculation, but are ultimately eliminated. More specifically, we find the following:

\begin{itemize}
	\item The stress tensor in~\eqref{deflinres} is off shell with respect to gravity, in the sense that the background dynamics has not yet been fixed~\cite{Ota:2025rll}. Once the background is determined, gravity couples to matter, and an additional non-dynamical local term must be included to obtain \eqref{eomschem}. We refer to this as the on-shell projection~$T^{\rm os}_{\mu\nu}$, and we have
	\begin{align}
		T^{\rm lin.}_{\mu\nu} = T_{\mu\nu} - T^{\rm bg}_{\mu\nu}+T^{\rm os}_{\mu\nu} = T^{\rm os}_{\mu\nu} + T^{\rm dyn.}_{\mu\nu}+ T^{\rm cont.}_{\mu\nu}.
	\end{align} 
	Thus, it is not enough to linearize $T_{\mu\nu}$ by subtracting $T^{\rm bg}_{\mu\nu}$. This is the usual situation in cosmological perturbation theory, but it was not explicit in thermal field theory in a Minkowski background, where the thermal Minkowski limit is on shell by construction.

	\item The genuine non-dynamical response is the sum of the on-shell projection~$T^{\rm os}_{\mu\nu}$ and the contact response~$T^{\rm cont.}_{\mu\nu}$. The net on-shell projection is spin-independent, whereas $T^{\rm cont.}_{\mu\nu}$ generally depends on the action, and hence on the spin. It is found that both $T^{\rm os}_{\mu\nu}$ and $T^{\rm cont.}_{\mu\nu}$ depend on the expansion scheme of the metric tensor, while their sum is field-redefinition independent.
	 
	\item The dynamical response $T^{\rm dyn.}_{\mu\nu}$ contains two time scales: the inverse temperature $\beta \equiv 1/T$, and the cosmological time scale $\tau$:
	\begin{align}
		T^{\rm dyn.}_{\mu\nu} = T^{\rm fast}_{\mu\nu} + T^{\rm slow}_{\mu\nu}.
	\end{align} 
	The component varying on the short scale $\beta \ll \tau$ is the fast response $T^{\rm fast}_{\mu\nu}$ and behaves as a local term on cosmological scales. This fast response contains a nontrivial spin dependence that must be distinguished from $T^{\rm os}_{\mu\nu}+ T^{\rm cont.}_{\mu\nu}$. The slow nonlocal response $T^{\rm slow}_{\mu\nu}$ is found to be consistent with the kinetic-theory prediction.
	\item For spins 0, 1/2, and 1, the non-dynamical local term cancels the fast response:
	\begin{align}
		T^{\rm os}_{\mu\nu} + 	T^{\rm fast}_{\mu\nu} + 	T^{\rm cont.}_{\mu\nu} =0.
	\end{align}
	This cancellation is not trivial, as their origins are different: $T^{\rm os}_{\mu\nu}  + 	T^{\rm cont.}_{\mu\nu}$ arises from the genuine local response, while $T^{\rm fast}_{\mu\nu}$ is generated by the nonlocal commutator. 
    At zero temperature, the integral suffers from the usual UV divergence. After renormalization, the spin-dependent contributions may be absorbed into a redefinition of physical constants, and the remaining dynamical part becomes spin independent. In the thermal fields setup considered here, however, such UV divergences are absent because of the exponential suppression provided by the thermal distribution. This leaves no freedom to absorb the spin-dependent coefficients. Remarkably, through explicit and careful calculations, we find that these potentially problematic local terms cancel exactly within linear response theory.

	\item The spin-dependent coefficients \eqref{cft} appear in the IR limit of the thermal expectation value of the commutator 
	\begin{align}
		\lim_{k\to 0} T^{\rm dyn.}_{ij} \propto (n_0 \rho_b  - 3n_{\frac{1}{2}} \rho_f +12 n_1 \rho_b),
	\end{align}
	where $\rho_b$ and $\rho_f$ are the energy densities of relativistic bosonic and fermionic degrees of freedom.
	The overall weights are $\rho_b:-3\rho_f:12 \rho_b$.
	The minus sign in the spin-${1}/{2}$ contribution is due to the thermal average. This sign is reversed for the vacuum-energy contribution we dropped, which is consistent with the zero-temperature calculation. The spin-dependent coefficient does not manifest in the final effective dynamics of gravitational waves, and observations are sensitive to the slow part only.

	\item The one-loop analysis of the graviton two-point function~\cite{Ota:2023iyh,Frob:2025sfq,Ota:2024idm} is essentially equivalent. However, the present approach allows us to obtain the thermal effective action for gravitational perturbations, so that the resummed spectrum follows. Our equation of motion is linearized, with the four-point vertex implicit in \eqref{deflinres}. In the end, the single-vertex diagram in the one-loop analysis corresponds to the non-dynamical response once the on-shell projection and the contact term are combined. We have verified this explicitly for spins $0$, $1/2$, and $1$. 
\end{itemize}

These results are summarized in Table~\ref{tab:summary:full}.
In the following, we give a detailed derivation of the above results.

\renewcommand{\arraystretch}{1.5}
\begin{table}
\centering
\begin{tabular}{|c|c|c|c|c|c|c|}
\hline
\rowcolor{gray!20}
\textbf{Spin}  & \textbf{Slow}~$(A)$ &\textbf{Fast}~$(B)$ & \textbf{Contact}~$(C)$ & \textbf{On-shell}~$(D)$ & $A+B$ & $B+C+D$   \\ \hline\hline
\rowcolor{gray!8} 0 &  $+\frac{8}{5}$~\eqref{eq:pi:sp:0}   & $-2$~\eqref{qisc} & 0~\eqref{ctminisc} & $+2~\eqref{osterm}$ &  $-\frac{2}{5}$  & $0$  \\ \hline
$\frac{1}{2}$ & $+\frac{16}{5}$~\eqref{eq:pi:sp:1/2} &$-2$~\eqref{qispin1/2} & $-2$~\eqref{cont1/2} & $+4$~\eqref{osterm} & $+\frac{6}{5}$ & $0$ \\ \hline
\rowcolor{gray!8} 1 & $+\frac{16}{5}$~\eqref{eq:pi:sp:1} &$-8$~\eqref{qispin1} & $+4$~\eqref{contspint1} & $+4$~\eqref{osterm} & $-\frac{24}{5}$ & $0$ \\ \hline
\end{tabular}
\caption{Summary of the local contributions for different spins in units of $\frac{\kappa^2\rho_{b/f}}{12}$ with $\rho_{b/f}$ being a relativistic bosonic/fermionic energy density (with $k\to 0$ for ``Slow''). For comparison, the entries are normalized to the mass shift in the effective equation of motion for the gravitational wave~\eqref{effectiveEOM}. $A$ gives the mass shift in the dispersion relation, whereas this same mass shift is subtracted so that it vanishes in the real-time integral kernel in~\eqref{effectiveEOM}. The combination $A+B$ corresponds to the IR correction for unequal-time stress tensor correlators, whose ratio is $1:-3:12$. Although the fast response~$B$, the contact contribution~$C$, and the on-shell contribution~$D$ depend on the spin, they cancel exactly in the sum, yielding a spin-independent equation of motion~\eqref{effectiveEOM}. $A$, $B$, and $C+D$ are independent of the field redefinition of the gravitational perturbations.}
\label{tab:summary:full}
\end{table}

\section{Common structure}
Before turning to the detailed analysis for each spin, we summarize several technical structures common to the present analysis.

\subsection{Conventions}
Hereafter we restrict our attention to gravitational waves, so in addition to the convention~\eqref{defmetric}, we work in the transverse traceless~(TT) gauge:
\begin{align}
	h^{0\mu} = \delta_{ij} h^{ij} = \partial_i h^{ij} = 0.
\end{align}
Latin indices are raised and lowered with the flat-space metric.
The spatial indices of external gravitational-wave momenta are always projected to the TT subspace unless otherwise stated.
This gauge choice simplifies various technical aspects.
The treatment of general cosmological perturbations is left for future work.
Chiral fermions are introduced following the conventions of Martin~\cite{Martin:1997ns}.
For the Maxwell field, we use the Coulomb gauge.
The Minkowski metric is $\eta_{\mu\nu}={\rm diag}(-1,1,1,1)$.
As the hard loops are provided by conformal fields, we may compute the response in the Minkowski background and obtain the result in the FLRW background by Weyl rescaling
\begin{align}
	T_{\mu\nu} \to a^{-2} T_{\mu\nu}. 
    \label{Eq.ToFLRW}
\end{align}
We Fourier transform only in the three spatial directions, since the FLRW background is time dependent
\begin{align}
	\mathcal O(x^0,\mathbf{k}) = \int d^3 x\,e^{-i\mathbf{k}\cdot \mathbf{x}}\mathcal O(x).
\end{align}
Gravitational waves in real space are expanded as
\begin{align}
	h_{ij}(\tau,\mathbf{x}) = \int \frac{d^3k}{(2\pi)^3} e^{i\mathbf{k} \cdot \mathbf{x}}h_{ij}(\tau,\mathbf{k}).
\end{align}
The Fourier modes are further decomposed in terms of the gravitational polarization tensors:
\begin{align}
	h_{ij}(\tau,\mathbf{k}) = \sum_{s=\pm}e^{(s)}_{ij}(\mathbf{k})h^{(s)}(\tau,\mathbf{k}).
\end{align}
where $s$ is the labels for helicity and the transverse-traceless polarization tensors satisfy
the following properties
\begin{align}
	e^{(s)}_{ij}(\mathbf{k}) \left[e^{(s')ij}(\mathbf{k})\right]^*&= \delta^{ss'}\,,\\
    \left[e^{(s)}_{ij}(\mathbf{k})\right]^*&=e^{(-s)}_{ij}(\mathbf{k})=e^{(s)}_{ij}(\mathbf{-k})\,. 
\end{align}
One may freely choose $\mathbf{k}$ to lie along the $z$-axis, and parameterize the hard three-momentum as 
\begin{align}
	p_i = \left(p \sqrt{1-\mu^2}\cos \varphi,p \sqrt{1-\mu^2}\sin \varphi , p\mu \right)\,,
\end{align}
where $\mu\equiv\hat{\mathbf{k}}\cdot\hat{\mathbf{p}}=\cos 
\theta$ and $\varphi$ is the azimuthal angle. A useful spin algebra identity is, for example,
\begin{align}
	\left[e^{(s)ij}(\mathbf{k})\right]^*p_i p_j p_k p_l\,h^{kl}(\tau,\mathbf{k}) 
	&= \sum_{s'=\pm} \left[e^{(s)ij}(\mathbf{k})\right]^* p_i p_j p_k p_l  e^{(s')kl}(\mathbf{k}) h^{(s')}(\tau,\mathbf{k})
	\notag
	\\
	&=\frac{p^4}{4}(1-\mu^2)^2 \sum_{s'=\pm} e^{i(s'-s)\varphi} h^{(s')}(\tau,\mathbf{k}).\label{sponalg1}
\end{align}
Similarly, we will also use the identity
\begin{align}
	\delta_{jl} \left[e^{(s)ij}(\mathbf{k})\right]^*p_i  p_k   h^{kl}(\tau,\mathbf{k}) 
	&=\frac{p^2}{2}(1-\mu^2) h^{(s)}(\tau,\mathbf{k}).\label{sponalg2}
\end{align}

\subsection{On-shell projection}
The structure of the on-shell projection is common to all spins.
In the TT gauge, the Einstein tensor and the stress tensor are written as~\cite{Weinberg:2008zzc,Mukhanov:2005sc} 
\begin{align}
	G_{00} &= 3\mathcal H^2, \\
	G_{ij} &= - \frac{\kappa}{2} (h_{ij}'' + 2 \mathcal H h_{ij}' - \nabla^2 h_{ij}) - \left( \mathcal H^2 + 2 \mathcal H' \right) (\delta_{ij} - \kappa h_{ij})\,,
\end{align}
with the conformal Hubble $\mathcal{H}\equiv a'/a$. The leading-order stress tensor is
\begin{align}
		T^{\rm bg}_{00} & = a^2 \rho, \\
	T^{\rm bg}_{ij} &= a^2 P \delta_{ij}.
\end{align}
The Einstein equation
\begin{align}
	G_{\mu\nu} = \frac{\kappa^2}{4}T_{\mu\nu},
\end{align}
is satisfied order by order in cosmological perturbations. Then the background equations of motion are given by
\begin{align}
	3\mathcal H^2 = \frac{\kappa^2}{4}a^2 \rho, \quad \mathcal H^2 + 2 \mathcal H' + \frac{\kappa^2}{4} a^2 P = 0.
\end{align}
The on-shell Einstein tensor for the gravitational-wave part is now expressed as 
\begin{align}
		G_{ij} &= - \frac{\kappa}{2} (h_{ij}'' + 2 \mathcal H h_{ij}' - \nabla^2 h_{ij}) -  \frac{\kappa^2}{4} a^2 P  \kappa h_{ij},
\end{align}
and we can read off
\begin{align}
	T^{\rm os}_{ij} = a^2 P  \kappa h_{ij}.\label{osterm}
\end{align}
In other words, the linearized Einstein equation can be written as
\begin{align}
	h_{ij}'' + 2 \mathcal H h_{ij}' - \nabla^2 h_{ij} =\pi_{ij},~\pi_{ij}\equiv  -\frac{2}{\kappa} \frac{\kappa^2}{4}T^{\rm lin}_{ij}. \label{hEoM}
\end{align}
That is, the first-order stress tensor is projected to 
\begin{align}
	T_{ij} \to T^{\rm lin}_{ij} = T^{\rm os}_{ij} +  T^{\rm dyn.}_{ij}+  T^{\rm cont.}_{ij}   .\label{oshellproject}
\end{align}
We refer to this procedure as the on-shell projection, which is common in cosmological perturbation theory, but has not been emphasized in the context of the graviton HTL analysis in the Minkowski limit.

\subsection{General structure of the response}

We must determine the anisotropic stress $\pi_{ij}$ from linear response theory, but the perturbed component in the dynamical response cannot be identified straightforwardly with $\pi_{ij}$.
The identification proceeds in the following steps:
\begin{enumerate}
	\item Find the off-shell linear response $T^{\rm dyn.}_{ij}+  T^{\rm cont.}_{ij}$.
	\item Separate the dynamical contributions into fast and slow responses, $T^{\rm dyn.}_{ij} = T^{\rm fast}_{ij} +T^{\rm slow}_{ij}$.
	\item Solve for the background spacetime and obtain $T^{\rm os}_{ij}$.
	\item Compute the anisotropic stress $\pi_{ij}$ from $T^{\rm lin}_{ij} = T^{\rm os}_{ij} +  T^{\rm dyn.}_{ij}+  T^{\rm cont.}_{ij}$.
	\item Refine the time boundary so as to respect the large diffeomorphism symmetry for IR graviton modes.
\end{enumerate}

\subsection{Green functions}

The retarded and advanced Green functions for the hard particles are commonly expressed in terms of the Pauli-Jordan function
\begin{align}
	G(x-y) &= -\int \frac{d^3 p}{(2\pi)^3} \frac{\sin\left(p_\mu(x^\mu-y^\mu)\right)}{p}
	\notag 
	\\
	&= \int \frac{d^3 p}{(2\pi)^3}e^{i\mathbf{p} \cdot (\mathbf{x} -\mathbf{y}) } \frac{\sin\left(p(x^0-y^0)\right)}{p}.
\end{align}
Using this function, the retarded Green function of a conformal scalar $\chi$ is
\begin{align}
	G^R(x-y) &= i\Theta(x^0-y^0)[\chi(x),\chi(y)]	= \Theta(x^0-y^0) G(x-y) ~ , \label{GRscalar}
\end{align}
where $\Theta$ is the Heaviside step function. The retarded and advanced Green functions of a Weyl fermion $\psi_{\alpha}$ are
\begin{align}
	G^R_{\alpha \dot \beta}(x-y) &=-i \Theta(x^0-y^0) \{ \psi_\alpha(x) ,\psi^\dagger_{\dot \beta} (y) \} = -i \Theta(x^0-y^0)\sigma^\mu_{\alpha \dot \beta} \partial^{(x)}_\mu G(x-y) ,
\\
	G^A_{\alpha \dot \beta}(x-y) &=i \Theta(y^0 - x^0) \{ \psi_\alpha(x) ,\psi^\dagger_{\dot \beta} (y) \} =  i \Theta(y^0 - x^0)\sigma^\mu_{\alpha \dot \beta} \partial^{(x)}_\mu G(x-y).
\end{align}
This implies symmetry under exchanging $x$ and $y$:
\begin{align}
	\{ \psi_\alpha(y) ,\psi^\dagger_{\dot \beta} (x) \} = \sigma^\mu_{\alpha \dot \beta} \partial^{(y)}_\mu G(y-x)= \sigma^\mu_{\alpha \dot \beta} \partial^{(x)}_\mu G(x-y)=\{ \psi_\alpha(x) ,\psi^\dagger_{\dot \beta} (y) \}.\label{27eq}
\end{align}
For a Maxwell field $A_\mu$, we have 
\begin{align}
G^R_{\rho\nu}(x-y)=		 i \Theta(x^0-y^0) [A_\rho(x), A_{\nu}(y)]&= \Theta(x^0-y^0) P_{\rho \nu} G(x-y),\label{GRA}
\end{align}
with the projection tensor in Coulomb gauge 
\begin{align}
	P_{ij} =  \delta_{ij} - \frac{\partial_i \partial_j}{\nabla^2},~P_{00}=P_{0i}=P_{i0}=0.
\end{align}

Let $b^\dagger_A(\mathbf{k})$ and $b_A(\mathbf{k})$ be creation and annihilation operators of a relativistic degree of freedom labeled by $A$. 
We define the thermal state by
\begin{align}
\langle  b^\dagger_{A}(\mathbf{p}) b_{B}(\mathbf{k})\rangle &=(2\pi)^3 \delta^{(3)}(\mathbf{k}-\mathbf{p})\delta_{AB}f_A,\label{eqdefd}
\end{align}
with the phase space distribution $f_A$, which may be bosonic or fermionic, and may describe particles or antiparticles for nonzero chemical potential.
Since a thermal ensemble is diagonal in the number basis, expectation values with an odd number of operators vanish. 
We eliminate the vacuum contribution by normal ordering.
Also, ${\rm Tr}[\hat D bb]={\rm Tr}[\hat D b^\dagger b^\dagger]=0$.
Eq.~\eqref{eqdefd} is equivalent to introducing the (grand) canonical ensemble in the Minkowski background.
For a Weyl fermion, $f_A=\{f,g\}$ are the Fermi-Dirac distributions of the particle and antiparticle:
\begin{align}
	f= \frac{1}{e^{\beta (k-\tilde \mu) }+1},~g= \frac{1}{e^{\beta(k+\tilde \mu)}+1}.\label{fddist}
\end{align}
Then the Keldysh propagator in this thermal state is expressed as
\begin{align}
	G^K (x-y)&= \int \frac{d^3 k}{(2\pi)^3} \frac{1}{2k}  \left( f(k)   e^{i k_\nu (x^\nu -y^\nu)} 
		+ g(k)   e^{-i k_\nu (x^\nu -y^\nu)} 
		\right)
		\notag 
		\\
&= \int \frac{d^3 k}{(2\pi)^3} e^{i\mathbf{k} \cdot (\mathbf{x} - \mathbf{y})}\frac{1}{k}\left(
\frac{f(k)+g(k)}{2} \cos k (x^0 -y^0) 
\right.
\nonumber\\
&
\left. +
\frac{f(k)-g(k)}{2i}\sin k (x^0 -y^0) \right).
\end{align}
Note that we drop the zero-temperature part.
For a small chemical potential $\tilde \mu$,
\begin{align}
f - g &\simeq -2 \tilde \mu \frac{\partial f}{\partial k},
\end{align}
and we can further simplify the above equation as
\begin{align}
	G^K (x-y)
&= \int \frac{d^3 k}{(2\pi)^3} e^{i\mathbf{k} \cdot (\mathbf{x} - \mathbf{y})}\frac{\cos k (x^0 -y^0)}{k}
f(k)
\notag 
\\
&+i\tilde \mu \int \frac{d^3 k}{(2\pi)^3} e^{i\mathbf{k} \cdot (\mathbf{x} - \mathbf{y})}\frac{\sin k (x^0 -y^0)}{k}
\frac{\partial f(k)}{\partial k}  .
\end{align}
Using $G^K$, one can express the Keldysh Green functions of free real scalar and vector fields as
\begin{align}
	\frac{1}{2}\left\langle \{ \chi(x) , \chi(y) \}\right\rangle & = G^K(x-y),\label{keldsca}
	\\
 	\frac{1}{2}\langle  \{A_\rho(x), A_{\nu}(y)\}\rangle  &=P_{\rho\nu} G^K(x-y).\label{GKA}
\end{align}
For a Weyl fermion, we have
\begin{align}
	 \langle \psi^\dagger_{\dot \beta} (y)  \psi_\alpha(x)\rangle
	= i\sigma^\mu_{\alpha \dot \beta} \partial^{(x)}_\mu G^K (x-y).
		\label{eqden}
\end{align}
For nonzero chemical potential, $G^K$ is complex.
We will focus on the zero chemical potential case ($\tilde \mu=0$) in this work.
Then
\begin{align}
	\langle \psi^\dagger_{\dot \beta} (x)  \psi_\alpha(y)\rangle
	= i\sigma^\mu_{\alpha \dot \beta} \partial^{(y)}_\mu G^K (y-x) =- i\sigma^\mu_{\alpha \dot \beta} \partial^{(x)}_\mu G^K (x-y) = - \langle \psi^\dagger_{\dot \beta} (y)  \psi_\alpha(x)\rangle.\label{218eq}
\end{align}

\section{Spin 0: conformal scalar}

We first examine the spin-0 case and carefully analyze the structure of the nonlocal kernel in the real-time formalism.
We identify a mismatch in the time boundary in the naive in-in formalism, propose a prescription to resolve it, and discuss the cancellation of local responses.

\subsection{Dynamical response}

Let us start from the action of a conformal scalar $\chi$:
\begin{align}
	 S_{\rm{scalar}}=-\frac{1}{2} \int d^4 x \sqrt{-g} \left( g^{\mu \nu}\partial_\mu \chi \partial_\nu \chi + \frac{1}{6}R\chi^2\right).
\end{align}
The stress tensor is then
\begin{align}
	\hat T_{\mu\nu} = \partial_\mu \chi \partial_\nu \chi  - \frac{1}{2}g_{\mu\nu}g^{\rho \sigma}\partial_\rho \chi \partial_\sigma \chi + \frac{1}{6}\left(g_{\mu\nu}g^{\rho \sigma}\nabla_\rho \nabla_\sigma - \nabla_\mu \nabla_\nu +G_{\mu\nu} \right)\chi^2.
\end{align}
The interaction-picture stress tensor in the limit $\kappa \to 0$ reduces to 
\begin{align}
	\hat T^I_{ij} = \partial_i \chi \partial_j \chi + \cdots, 
\end{align}
where $\cdots$ denotes the trace part.
The interaction Hamiltonian~\eqref{defHint} is
\begin{align}
	H_{\rm int} = \frac{\kappa}{2}\int d^3 y   h^{kl}(y)\partial_k\chi(y)\partial_l\chi(y),
\end{align}
and then the dynamical response is expressed as
\begin{align}
	T^{\rm dyn.}_{ij}(x)&=i\int^{x^0}dy^0 \left[H_{\rm int},\hat T^I_{ij}(x)\right] 
	\notag 
	\\
	&=i\frac{\kappa}{2}\int^{x^0}dy^0 \int d^3 y   [\partial_k\chi(y)\partial_l\chi(y),\partial_i \chi(x) \partial_j \chi(x)]h^{kl}(y) .
\end{align}
Note that this equality holds up to the TT projection for the $ij$ indices.
Thus, the graviton dynamical response for the conformal scalar is equivalent to that of a minimally coupled scalar. 
The commutator simplifies to
\begin{align}
	i\left[ \partial_k \chi(y) \partial_l \chi(y) ,\partial_i \chi(x) \partial_j \chi(x) \right] 
	&=2i\left[\chi(y) , \chi(x) \right] \partial^{(y)}_l \partial^{(y)}_k \partial^{(x)}_i  \partial^{(x)}_j \{ \chi(y) , \chi(x) \} . 
\end{align}
The transverse condition allows integration by parts with respect to the indices $i,j,k,l$.
Then, using the Green functions~\eqref{GRscalar} and \eqref{keldsca} 
\begin{align}
	T^{\rm dyn.}_{ij}(x) =-2\kappa\int^{x^0}dy^0 \int d^3 y \, G(x-y)\partial_i \partial_j \partial_k\partial_l G_K(x-y) h^{kl}(y) .\label{eq225unko}
\end{align}
In Fourier space, using the identity Eq.~\eqref{sponalg1},
\begin{align}
	T^{\rm dyn.(s)}(x^0,\mathbf{k}) &=- 2\kappa \int^{x^0}dy^0 
	 \int \frac{p^2 dp }{2\pi^2}\int \frac{d\mu}{2}
	 \notag 
	 \\
	 & \times p^4 \frac{\sin\left(p'(x^0-y^0)\right)}{p'}\frac{\cos\left(p(x^0-y^0)\right)}{p} f(p) 
	 \frac{(1-\mu^2)^2}{4}
	 h^{(s)}(y^0,\mathbf{k}),
\end{align}
where $p' \equiv |\mathbf{k} - \mathbf{p}|$.
We have found an analytic expression for this $(\mu,p)$ integral: 
\begin{align}
T^{\rm dyn.}_{ij}(x^0,\mathbf{k})& = - \kappa\frac{2}{15} \pi^{2} T^{4} \int^{x^0} dy^0 h_{ij}(y^0,\mathbf{k})\frac{\partial}{\partial y^0}K(k (x^0-y^0) )
	\notag
	\\
	& -\kappa T^4 \int^{Tx^0} dw h_{ij}\left(x^0-\frac{w}{T},\mathbf{k}\right) 
	\notag 
	\\
	& \times \left[(  A_{0}(w) +\left(\frac{k}{T}\right)^2w^2 A_{2}(w) )  K\left(\frac{k}{T} w\right) 
+  B_{0}(w) \frac{\partial}{\partial w}K\left(\frac{k}{T} w\right)
    \right],
\end{align}
with $w=T(x^0-y^0)$ and the explicit forms of these functions given by
\begin{align}
A_{0}(x)
&= -\frac{3}{8 \pi^{2} x^{5}} 
  {+}\pi^{3}  
   \bigl(11\cosh(2\pi x)+\cosh(6\pi x)\bigr) 
   \operatorname{csch}(2\pi x)^{5} 
\\
A_{2}(x)
&= 
 \frac{5 + 2\pi^{2} x^{2}}{4 \pi^{2} x^{5}} 
  {-} \frac{\pi }{8 \pi^{2} x^{4}} 
   \operatorname{csch}(2\pi x)^{3} 
   \Bigl[
     \bigl(-3 + 8\pi^{2} x^{2}\bigr)\cosh(2\pi x) \notag\\
&    + 3\bigl(\cosh(6\pi x) + 4\pi x \sinh(2\pi x)\bigr)
   \Bigr] 
\\
B_{0}(x)
&= 
  \frac{75}{8 \pi^{2} x^{4}}
  {+}\frac{7 }{2 x^{2}} 
  {-}\frac{1}{2\pi x^{3}} 
   \Bigl[
     21 \coth(2\pi x)+ 12\pi^{3} x^{3} 
       \operatorname{csch}(2\pi x)^{4} \notag\\
&     + \pi x 
       \bigl(21 + 8\pi^{2} x^{2}
       + 18\pi x \coth(2\pi x)\bigr) 
       \operatorname{csch}(2\pi x)^{2} 
   \Bigr] ,
   \\
	K(x)  &= -\frac{\sin x}{x^3} - \frac{3 \cos x}{x^4} + \frac{3 \sin x}{x^5}.\label{dykernel}
\end{align}

\subsection{Fast and slow response}

The nonlocal response convolved with the retarded kernel~\eqref{dykernel} is understood as the slow dynamical response on the time scale $\tau$.
By contrast, the terms convolved with $A_0$, $A_2$, and $B_0$ show a distinct behavior.
$A_0$, $A_2$, and $B_0$ act as window functions that are nonvanishing only for $w\lesssim 1$.
For $w\gg 1$,
\begin{align}
	A_0,A_2,B_0\to 0.
\end{align}
We also take the high-temperature limit, $k/T\to 0$.
Since $Tx^0\gg w$, we can approximate the $A_0$ term as
\begin{align} 
 \kappa \frac{T^4}{15} h_{ij}(x^0,\mathbf{k}) \int^{\infty} dw \, A_{0}(w)=  \kappa\,h_{ij}(x^0,\mathbf{k}) \frac{\pi^2}{450}T^4.
\end{align}
This is the fast response, in the sense that the dominant contribution arises from the short time scale $\beta = T^{-1}$, and we have used the property of the memory kernel in the IR limit $\lim\limits_{x\to 0} K(x)=1/15$.
Similarly, we obtain 
\begin{align}
 \kappa h_{ij}(x^0,\mathbf{k})\frac{T^4}{15}\left(\frac{k}{T}\right)^2 \int^{\infty} dw \, w^2A_{2}(w)=  T^4\mathcal O(k^2/T^2).
\end{align}
We formally extend the upper limit of integration to infinity in order to estimate the overall dependence on $k/T$. Nevertheless, the integral with respect to the dimensionless parameter $w$ exhibits a logarithmic divergence at infinity. In principle, the upper cutoff should therefore be at most $Tx_0$, which may introduce an additional $\log (T/H)$ factor. 
In any case, this term drops out in the high-temperature limit.
As for the remaining term,
\begin{align}
	\kappa  T^4 h_{ij}(x^0,\mathbf{k})
\int^{\infty} dw B_{0}(w) \frac{\partial}{\partial w} K\left(\frac{k}{T}w\right)
= T^4 \mathcal O(k^2/ T^2).
\end{align}
In summary, in the high-temperature limit, and restoring the scale factor by Weyl rescaling, we find the dynamical response in the stress tensor as
\begin{align}
	T^{\rm dyn.}_{ij}(x^0,\mathbf{k})= - \frac{\kappa}{15}a^2 \rho_b h_{ij}(x^0,\mathbf{k})  -4\kappa a^2\rho_b \int^{x^0} dy^0 h_{ij}(y^0,\mathbf{k})\frac{\partial}{\partial y^0}K(k (x^0-y^0) ),\label{spin0result}
\end{align}
where the energy density of a bosonic relativistic degree of freedom is
\begin{align}
	\rho_b = \frac{\pi^2T^4}{30 a^4}.
\end{align}
The coefficient of $h_{ij}$ in the $k\to 0$ limit is 
\begin{align}
-\frac{T^4}{15a^4} \int^{\infty} dw A_{0}(w)= -\frac{\rho_b}{15}.
\end{align}
Note that $T$ is understood as the comoving temperature after rescaling by $a$.
When placing this term on the left-hand side of the linearized Einstein equation \eqref{hEoM}, we multiply by $2/M_{\rm pl}^2$, which results in the tachyonic mass $m_{\rm eff}^2 =-2H^2/5$ found in previous work~\cite{Ota:2023iyh,Frob:2025sfq,Ota:2024idm,Ota:2025yeu}.

This $m_{\rm eff}^2$ is the plasmon-like mass shift introduced by the dynamical response: it is a mass in the sense of the dispersion relation.
In the IR limit, a constant source arises from the boundary in the kernel integral, and a part of $m_{\rm eff}^2$ is subtracted. 
Hence, the non-vanishing local component in the IR limit is obtained after integrating by parts in Eq.~\eqref{spin0result}:
\begin{align}
	T^{\rm dyn.}_{ij}(x^0,\mathbf{k})  = & -\frac{1}{3} \kappa a^2\rho_b h_{ij}(x^0,\mathbf{k}) +4\kappa a^2\rho_b K(kx^0)h_{ij}(0,\mathbf{k})
	 \notag 
	 \\
	 &+ 4\kappa a^2\rho_b \int^{x^0} dy^0K(k (x^0-y^0) ) \frac{\partial}{\partial y^0}h_{ij}(y^0,\mathbf{k}).\label{nonlocsc}
\end{align}
In this form, the kernel integral vanishes in the IR region.
One can read off the fast response from \eqref{nonlocsc} and find
\begin{align}
	T^{\rm fast}_{ij} = -\frac{1}{3} \kappa a^2\rho_b h_{ij}(x^0,\mathbf{k}),\label{qisc}
\end{align}
which looks local but originates from a nonlocal kernel.
The integral term can be interpreted as the inhomogeneous solution of the nonlocal self-energy operator.
The fast response corresponds to the response in a very short time interval $T(x^0-y^0)\sim 1$, which can be seen explicitly in our real-time formalism.
In kinetic theory or in the imaginary-time formalism, the fast response is indistinguishable from the genuine local responses.

\subsection{Contact term for scalar}

As the last contributions, we evaluate $T^{\rm cont.}_{\mu\nu}$ for a scalar field.
For simplicity, let us first consider the minimally coupled scalar case.
Write
\begin{align}
	J_{\mu\nu} = \partial_\mu \chi \partial_\nu \chi,~J = g^{\mu\nu} J_{\mu\nu}.
\end{align}
Then the stress tensor operator $\hat T_{\mu\nu}$ can be rewritten as
\begin{align}
	\hat T_{\mu\nu} = J_{\mu\nu} - \frac{1}{2} g_{\mu\nu} J.
\end{align}
The variation of the stress tensor is
\begin{align*}
	\int d^4 x \frac{\delta \hat T_{\mu\nu}}{\delta g^{\rho \sigma}(x)} = - \frac{1}{2} \int d^4 x \left[  \frac{\delta g_{\mu\nu}}{\delta g^{\rho\sigma}(x)} g^{\alpha \beta} J_{\alpha \beta} + g_{\mu\nu} \frac{\delta g^{\alpha \beta}}{\delta g^{\rho\sigma}(x)} J_{\alpha \beta} \right].
\end{align*}
Then, together with
\begin{align}
	\frac{\delta g_{\mu\nu}(y)}{\delta g^{\rho \sigma}(x)} &=- \delta^{(4)}(x-y) \frac{1}{2}(g_{\mu\rho}g_{\nu\sigma} + g_{\nu\rho}g_{\mu\sigma}) 
\\
	\frac{\delta g^{\alpha \beta}(y)}{\delta g^{\rho \sigma}(x)} &= \delta^{(4)}(x-y)\frac{1}{2} \left(\delta^\alpha_{\rho} \delta^\beta_{\sigma} + \delta^\alpha_{\sigma} \delta^\beta_{\rho}\right),
\end{align}
and the ensemble average
\begin{align}
	\langle J \rangle_{\kappa =0} = 0,~\langle J_{\mu\nu} \rangle_{\kappa =0} = T^{\rm bg}_{\mu\nu},
\end{align}
with the subscript indicating $\kappa =0$, 
we have 
\begin{align}
	\int d^4 x \left\langle \frac{\delta \hat T_{\mu\nu}}{\delta g^{\rho\sigma}(x)}\right\rangle_{\kappa =0} \delta g^{\rho \sigma}(x) = - \frac{1}{2}\bar g_{\mu\nu}T^{\rm bg}_{\rho \sigma} \delta g^{\rho \sigma} = 0,
\end{align}
where the last equality holds for the TT tensor. 
Since the stress tensor is traceless, this term does not generate a contact term for the graviton. 
In summary,
\begin{align}
	T^{\rm cont.}_{\mu\nu} = 0,\label{ctminisc}
\end{align}
for a minimally coupled scalar.
For a conformal scalar, we have the additional coupling
\begin{align}
	-\frac{1}{12} \int d^4x \sqrt{-g} R\chi^2. 
\end{align}
This implies that the particle trajectories are no longer geodesic, and a deviation from standard kinetic theory is expected.
Then, the correction to the minimally coupled scalar stress tensor is 
\begin{align}
	\frac{1}{6}\left( G_{\mu\nu}  + g_{\mu\nu} \Box - \nabla_{\mu} \nabla_\nu   \right)\chi^2.
\end{align}
We can further take the variation with respect to $g^{\rho \sigma}$ to find the correction to the contact term in a general form, but we can show that this correction vanishes in the high-temperature limit.
Since the derivative acting on $\chi^2$ is a total derivative that reduces to the external momentum, the thermal average over $\chi^2$ yields $T^2$.
Then, the geometric coefficient can no longer provide an additional power of $T$.
Hence, the correction from the conformal coupling is either $\mathcal O(k^2 T^2)$ or $\mathcal O(H^2 T^2)$.
These are suppressed by factors of $k/T$ and $H/M_{\rm pl}$, which are ignored in the present analysis.
Thus, in the HTL limit (with $H/M_{\rm pl} \ll 1$), the minimally coupled scalar and the conformal scalar yield the same result.

\subsection{Cancellation of local responses}

From Eqs.~\eqref{oshellproject},~\eqref{qisc}, and~\eqref{ctminisc}, one can eliminate the scale-independent local contribution from the linear response:
\begin{align}
	T^{\rm os}_{ij} +T^{\rm fast}_{ij}+T^{\rm cont.}_{ij} = 0. 
\end{align}
The scale-dependent source in the dynamical response (i.e. the second term in Eq.~\eqref{nonlocsc}) can be eliminated by using the ambiguity in the shift of the homogeneous solution~\cite{Rebhan:1994zw,Ota:2025rll}
\begin{align}
	a^2\rho_b \epsilon_{ij}(\mathbf{k})K(kx).\label{bcextra}
\end{align}
This point is also discussed in appendix~\ref{appkernel}.
Using this freedom, one may eliminate the secular growth term in the $kx\to 0$ limit.
This suggests a mismatch in the time boundary for the stress-tensor retarded propagator in the standard in-in formalism.
Our linear response function is constructed from the free UV propagators $G^R$ and $G^K$ represented in $(\tau, \mathbf{k})$ space, whose initial data are specified in flat spacetime.
The linear response problem for the stress tensor is essentially a problem of inverting the self-energy.
The initial data for this problem are not equivalent to those of flat spacetime, and the spacetime is asymptotically anisotropic due to the IR gravitational waves.
We must correct the time boundary for this reason.

\subsection{Summary of spin 0}
Combining Eqs.~\eqref{osterm}, \eqref{nonlocsc}, and \eqref{ctminisc}, the shear tensor is found to be
\begin{align}
	\pi_{ij} = -24 \kappa a^2 \left(\frac{\kappa^2}{12}\rho_b\right) \int^{x^0} dy^0K(k (x^0-y^0) ) \frac{\partial}{\partial y^0}h_{ij}(y^0,\mathbf{k}).\label{eq:pi:sp:0}
\end{align}
Under the Friedmann equation for a single scalar, radiation-dominated universe,
\begin{align}
	\frac{\kappa^2}{12}\rho_b = H^2,
\end{align}
this reproduces the kinetic-theory result.

\section{Spin $\frac{1}{2}$: Weyl spinor}

We follow the same steps for a left-handed Weyl fermion $\psi$.
Our convention for the spinor algebra follows Ref.~\cite{Martin:1997ns}.

\subsection{Dynamical response}
The interaction-picture stress tensor for a Weyl fermion is 
\begin{align}
	\hat T^I_{ij} = - i \psi^\dagger \bar \sigma_j \partial_i \psi + \cdots, 
\end{align}
with the trace part denoted by $\cdots$.
Note that the symmetrizing the indices and derivatives does not affect the final result.
The interaction Hamiltonian \eqref{defHint} is
\begin{align}
	H_{\rm int} = - i \frac{\kappa}{2} \int d^3x   h^{ij}\psi^\dagger \bar \sigma_j \partial_i \psi .
\end{align}
The dynamical response is therefore 
\begin{align}
		T^{\rm dyn.}_{ij}(x)
	&=-i\frac{\kappa}{2}\bar \sigma_l^{\dot \alpha \alpha}\bar \sigma_j^{\dot \beta \beta} \int^{x^0}dy^0 \int d^3 y    \left[ \psi^\dagger_{\dot \alpha}(y) \partial_k \psi_\alpha(y) ,  \psi^\dagger_{\dot \beta}(x)  \partial_i \psi_\beta(x) \right]h^{kl}(y) \label{unkoo}.
\end{align}
Grassmann-odd operators $A,B,C,D$ satisfy
\begin{align}	
	[AB,CD] 	&=  A \{B, C\} D -\{A, C\} BD  + CA  \{ B, D \} - C\{ A, D\}B . 
\end{align}
Using this with \eqref{27eq}, \eqref{218eq}, and \eqref{unkoo}, we get
\begin{align}
	\frac{\kappa}{2}\bar \sigma_l^{\dot \alpha \alpha}\sigma^\mu_{\alpha \dot \beta} \bar \sigma_j^{\dot \beta \beta}\sigma^\nu_{\beta \dot \alpha} \int^{x^0}dy^0 \int d^3 y   
	h^{kl}(y)
	 \partial^{(x)}_\mu G(x-y) \partial^{(x)}_k \partial^{(x)}_i  \partial^{(x)}_\nu G^K (x-y)+\mu \leftrightarrow \nu.\label{chemicalptnonzero?}
\end{align}
We use the following spinor-algebra identity ($\epsilon^{0123}=-1$)~\cite{Martin:1997ns}
\begin{align}
	{\rm Tr}[\bar \sigma^\mu \sigma_j \sigma^\nu \bar \sigma_l ] = 2 \delta_j{}^\mu\delta_l{}^\nu + 2 \delta_l{}^\mu\delta_j{}^\nu - 2 \eta^{\mu\nu}\delta_{lj} -2 i\epsilon^\mu{}_j{}^\nu{}_l. 
\end{align}
The antisymmetric part vanishes for the zero chemical potential case.
Then we get
\begin{align}
		T^{\rm dyn.}_{ij}(x) = A_{ij} + B_{ij},
\end{align}
with
\begin{align}
	A_{ij}=	&- 4 \kappa \int^{x^0}dy^0 \int d^3 y   
	h^{kl}(y)
	  G(x-y) \partial^{(x)}_j \partial^{(x)}_k \partial^{(x)}_i  \partial^{(x)}_l G^K (x-y)
\label{defAij}
	  \\
B_{ij}	   = &- 2\kappa \eta^{\mu\nu}\delta_{lj} \int^{x^0}dy^0 \int d^3 y   
	h^{kl}(y)
	 \partial^{(x)}_\mu G(x-y) \partial^{(x)}_k \partial^{(x)}_i  \partial^{(x)}_\nu G^K (x-y).
\end{align}
Note that $A_{ij}$ is exactly equal to two times the scalar case~\eqref{eq225unko}, up to the fermion phase-space distribution.
The factor of two is understood as the number of on-shell channels. 
Therefore, the kernel structure follows the scalar case for $A_{ij}$.
$B_{ij}$ is further decomposed into temporal-derivative and spatial-derivative parts
\begin{align}
	B_{ij} = B^T_{ij} + B^S_{ij},
\end{align}
with
\begin{align}
	B^T_{ij}	   = &	 2\kappa  \delta_{lj} \int^{x^0}dy^0 \int d^3 y   
	h^{kl}(y)
	 \partial^{(x)}_0 G(x-y) \partial^{(x)}_k \partial^{(x)}_i  \partial^{(x)}_0 G^K (x-y),
	 \\
	B^S_{ij}	   = &	 -2\kappa \delta^{mn}\delta_{lj} \int^{x^0}dy^0 \int d^3 y   
	h^{kl}(y)
	 \partial^{(x)}_m G(x-y) \partial^{(x)}_k \partial^{(x)}_i  \partial^{(x)}_n G^K (x-y).
\end{align}
We cannot integrate by parts anymore, since $\mu$ and $\nu$ are not contracted with the polarization tensor.
Using Eq.~\eqref{sponalg2}, the Fourier transforms of $B^T_{ij}$ and $B^S_{ij}$ are simplified as
\begin{align}
		e^{(s)ij}(-\mathbf{k})\,B^T_{ij} &=	\kappa  
 \int^{x^0}dy^0 h^{(s)}(y^0,\mathbf{k}) \int \frac{p^4dp}{2\pi^2} \int \frac{d\mu}{2} \notag 
 \\
 &\times 
  (1-\mu^2)  
    \cos\left(p'(x^0-y^0)\right)    
		\sin\left(p(x^0- y^0)\right) f(p),
\\
		e^{(s)ij}(-\mathbf{k})\,B^S_{ij} &=	\kappa  
 \int^{x^0}dy^0 h^{(s)}(y^0,\mathbf{k}) \int \frac{p^4dp}{2\pi^2} \int \frac{d\mu}{2} \notag 
 \\
 &\times 
 (1-\mu^2)\frac{p^2 - kp\mu}{pp'}  
    \sin\left(p' (x^0-y^0)\right)\cos\left(p (x^0-y^0)\right) f(p).
\end{align}
Just as in the scalar case, we have found the analytic integral over $(p,\mu)$.
In the high-temperature limit $k/T\to 0$, we have 
\begin{align}
		e^{(s)ij}(-\mathbf{k})\,B_{ij} = \kappa \frac{\rho_f}{3} h^{(s)}(x^0,\mathbf{k}) ,
\end{align}
where the energy density of a fermionic relativistic degree of freedom is
\begin{align}
	\rho_f = \frac{7}{8}\frac{\pi^2T^4}{30}.
\end{align}
Thus, $B_{ij}$ contains only the fast response, and the nonlocal response disappears in the HTL limit.
In summary, combining $A_{ij}$ and $B_{ij}$, we find
\begin{align}
	T^{\rm dyn.}_{ij}(x^0,\mathbf{k})= \frac{\kappa}{5}a^2 \rho_f h_{ij}(x^0,\mathbf{k})  -8\kappa a^2\rho_f \int^{x^0} dy^0 h_{ij}(y^0,\mathbf{k})\frac{\partial}{\partial y^0}K(k (x^0-y^0) ).\label{spin1/2result}
\end{align}
Comparing with Eq.~\eqref{spin0result}, although one might have expected the response function to be proportional to $C_T=3$, we find a factor of two instead.
$C_T$ appears instead in the IR dynamical response $\lim\limits_{k\to 0}T^{\rm dyn.}_{ij}$:
\begin{align}
	-\frac{\rho_b}{15}: \frac{\rho_f}{5} = \rho_b : -3\rho_f. 
\end{align}
The minus sign arises because $G^K$ changes sign for fermions at finite temperature while $G$ remains the same.
Although the zero-temperature contribution is divergent, the ratio of the spin-0 to spin-1/2 contributions becomes $1:3$. Interestingly, as we have repeatedly emphasized, although these $C_T$ factors will be canceled in the final results, the spin dependence does appear at this stage and exhibits a structure similar to that of the CFT two-point function in Eq.~\eqref{cft}; however, it is visible only in the limit $k\to0$. Very roughly speaking, the momentum $k$ of gravitational waves introduces a particular scale into the problem under consideration. Only in the IR limit do this scale dependence disappears, and we recover the CFT result in Eq.~\eqref{CTdef}. A detailed analysis of the deeper relations to other familiar cases of these spin-dependent coefficients will be left for future work.

Following the scalar case, Eq.~\eqref{spin1/2result} is recast as
\begin{align}
		T^{\rm dyn.}_{ij}(x^0,\mathbf{k})  &= -\frac{1}{3} \kappa a^2\rho_f h_{ij}(x^0,\mathbf{k}) +8\kappa a^2\rho_f K(kx^0)h_{ij}(0,\mathbf{k})
	 \notag 
	 \\
	 &+ 8\kappa a^2\rho_f \int^{x^0} dy^0K(k (x^0-y^0) ) \frac{\partial}{\partial y^0}h_{ij}(y^0,\mathbf{k}),\label{dyn1/2}
\end{align}
and then we identify the fast response as
\begin{align}
	T^{\rm fast}_{ij}(x^0,\mathbf{k})  = -\frac{1}{3} \kappa a^2 \rho_f h_{ij}(x^0,\mathbf{k}).\label{qispin1/2}
\end{align}

\subsection{Contact term for Weyl fermion}

The stress tensor operator for a Weyl fermion is
\begin{align}
	\hat T_{\mu\nu} =- i \psi^\dagger \bar \sigma^N E_N{}^\rho g_{\rho \nu} \nabla_\mu \psi +  g_{\mu\nu}\mathcal L,
\end{align}
up to symmetrization of indices and derivatives.
Capital indices are local Lorentz indices in the vierbein formalism.
Define
\begin{align}
	L_{\mu}{}^N = - i \psi^\dagger \bar \sigma^N \partial_\mu \psi.
\end{align}
Since the spin connection is second order in $h_{ij}$, we have
\begin{align}
	\mathcal L = L_{\mu}{}^NE_N{}^\mu + \mathcal O(\kappa^2).
\end{align}
Then the stress tensor is written as
\begin{align}
	\hat T_{\mu\nu} =   E_{\nu}{}^M  \eta_{MN} L_{\mu}{}^N +  E_N{}^\rho E_{\mu}{}^M E_{\nu}{}^L \eta_{ML} L_{\rho}{}^N  + \mathcal O(\kappa^2).
\end{align}
In flat spacetime with $h^{ij}$, the vierbein is  
\begin{align}
	E_0{}^{0} &= 1,
	\\
	E_0{}^{I} &= E_i{}^{0} = 0,
	\\
	E_i{}^I & = \left(\delta_i{}^j - \frac{1}{2} h_i{}^j \right)\delta_j{}^I,
	\\
	E_I{}^i & = \delta_I{}^j \left(\delta_j{}^i + \frac{1}{2} h_j{}^i \right).
\end{align}
The TT component of $T_{ij}$ is
\begin{align}
	\delta \hat T_{ij} =  - \frac{1}{2} h_{jk} L_{i}{}^k 
	+ \frac{1}{2} \delta_{ij} h_k{}^l  L_{l}{}^k
	-  h_{ij} L_{N}{}^N
	  + \mathcal O(\kappa^2).\label{equnko}
\end{align}
and therefore
\begin{align}
	\int d^4 x \left \langle \frac{\delta \hat T_{ij}}{\delta h^{mn}}\right\rangle_{\kappa=0} h^{mn}  = -
	  \frac{1}{2} \langle L_{i}{}^k \rangle h_{jk} = -\frac{P}{2}h_{ij}.
\end{align}
In summary, after the Weyl rescaling, we obtain
\begin{align}
	T^{\rm cont.}_{ij}= -a^2 P_fh_{ij}.\label{cont1/2}
\end{align}
Note that $P = 2 P_f$ as $P_f$ is the pressure of a relativistic fermionic degree of freedom.

\subsection{Summary of spin $\frac{1}{2}$}

Combining Eqs.~\eqref{osterm}, \eqref{dyn1/2}, and~\eqref{cont1/2}, the shear tensor from thermal fermionic field is found as
\begin{align}
	\pi_{ij} = -24 \kappa a^2 \left(\frac{\kappa^2}{6}\rho_f\right) \int^{x^0} dy^0K(k (x^0-y^0) ) \frac{\partial}{\partial y^0}h_{ij}(y^0,\mathbf{k}).\label{eq:pi:sp:1/2}
\end{align}
Replacing the Friedmann equation for a radiation-dominated universe with a single Weyl fermion
\begin{align}
	\frac{\kappa^2}{6}\rho_f = H^2,
\end{align}
this reproduces the kinetic-theory result.

\section{Spin 1: Maxwell field}

Finally, following the same steps as before, we move to the discussion of a Maxwell field.

\subsection{Dynamical response}
The field strength of the Maxwell field is defined as
\begin{align}
	F_{\mu\nu} = \partial_\mu A_\nu - \partial_\nu A_\mu,
\end{align}
and the stress tensor is given by
\begin{align}
	\hat T_{\mu\nu} = F_{\mu\rho}F_{\nu}{}^\rho - \frac{1}{4}g_{\mu\nu} F_{\rho \sigma} F^{\rho \sigma}.\label{def:emstresstensor}
\end{align}
In the Coulomb gauge, $A_0 = 0$ and $\partial_i A^i =0$.
The linear coupling~\eqref{defHint} is
\begin{align}
	H_{\rm int} = \frac{\kappa}{2}\int d^3 y   h^{kl}(y)F_{k}{}^\rho(y) F_{l \rho}(y).
\end{align}
We evaluate
\begin{align}
	i\int^{x^0}dy^0 [H_{\rm int},\hat T^I_{ij}(x)] 
	=\frac{i\kappa}{2}\int^{x^0}dy^0 \int d^3 y    \left[ F_{k}{}^\rho(y) F_{l \rho}(y) ,  F_{i}{}^\sigma(x) F_{j \sigma}(x) \right]h^{kl}(y) .
\end{align}
Using the symmetry under interchange of the dummy indices $kl$ and $ij$, the summation over these indices leads to
\begin{align}
	 \left[ F_{k}{}^\rho(y) F_{l \rho}(y) ,  F_{i}{}^\sigma(x) F_{j \sigma}(x) \right] = 4 \left[ F_{l\rho}(y),F_{j \sigma}(x)\right] \frac{F_{k}{}^{\rho}(y)F_{i}{}^{\sigma}(x) + F_{i}{}^{\sigma}(x) F_{k}{}^{\rho}(y)}{2}.  
\end{align}
Then applying Eqs.~\eqref{GRA} and \eqref{GKA}, we get
\begin{align}
	 \frac{\langle F_{k}{}^{\rho}(y)F_{i}{}^{\sigma}(x) + F_{i}{}^{\sigma}(x) F_{k}{}^{\rho}(y)\rangle }{2} 
	&= ( - \partial_k\partial_i P^{\rho \sigma}  + \partial^{\rho}\partial_i P_k{}^\sigma 
	+\partial_k\partial^{\sigma} P^\rho{}_i   - \partial^{\rho}\partial^{\sigma} P_{ki} ) G^K(x-y),
\end{align}
and also
\begin{align}
i\left[ F_{l\rho}(y),F_{j \sigma}(x)\right]
	&=
(-P_{\rho \sigma}\partial_l\partial_j -P_{jl}\partial_\rho\partial_\sigma + P_{\rho j}\partial_l\partial_\sigma + P_{l \sigma}\partial_\rho\partial_j)G(y-x).
\end{align}
We need to expand 
\begin{align}
-2\kappa \int^{x^0}dy^0 \int d^3 y &( - \partial_k\partial_i P^{\rho \sigma}  + \partial^{\rho}\partial_i P_k{}^\sigma 
	+\partial_k\partial^{\sigma} P^\rho{}_i   - \partial^{\rho}\partial^{\sigma} P_{ki} ) G^K(x-y)
	\notag 
	\\
	&\times (-P_{\rho \sigma}\partial_l\partial_j -P_{jl}\partial_\rho\partial_\sigma + P_{\rho j}\partial_l\partial_\sigma + P_{l \sigma}\partial_\rho\partial_j)G(x-y),
\end{align}
for finite $k$.
This algebraic calculation is tedious, but the dynamical response for the spin-1 case can be summarized as
\begin{align}
	T^{\rm dyn.}_{ij} = A_{ij} + B^{TT+SS}_{ij} + B^{TS}_{ij} ,
\end{align}
with the same local structure $A_{ij}$ defined in Eq.~\eqref{defAij}, and
\begin{align}
	e^{(s)ij}(-\mathbf{k})\,B^{TT+SS}_{ij} & =-\kappa  
 \int^{x^0}dy^0 h^{(s)}(y^0,\mathbf{k}) \int \frac{p^4dp}{2\pi^2} \int_{-1}^{1} d\mu 
 \notag 
 \\
 &\times \frac{p}{p'} \left(k^2 \left(\mu ^2+1\right)-2 k \mu  \left(\mu ^2+1\right) p+2 \mu ^2 p^2\right) \cos (p z) \sin \left(z p'\right)f(p)
	,
	\\ 
	e^{(s)ij}(-\mathbf{k})\,B^{TS}_{ij} & =2\kappa  
 \int^{x^0}dy^0 h^{(s)}(y^0,\mathbf{k}) \int \frac{p^2dp}{2\pi^2} \int_{-1}^{1} d\mu 
	    (k-p\mu) p \mu  \sin (pz)\cos (p'z)f(p), 
\end{align}
where $z\equiv x^0-y^0$. In the high-temperature limit, we find an analytic result:
\begin{align}
	\lim_{k\ll T}e^{(s)ij}(-\mathbf{k})\,(B^{TT+SS}_{ij} + B^{TS}_{ij}) & = -\frac{2}{3} \rho_b h^{(s)}(x^0,\mathbf{k}). 
\end{align}
In summary, we obtain
\begin{align}
	T^{\rm dyn.}_{ij}(x^0,\mathbf{k})= - \frac{12\kappa}{15}a^2 \rho_b h_{ij}(x^0,\mathbf{k})  -8\kappa a^2\rho_b \int^{x^0} dy^0 h_{ij}(y^0,\mathbf{k})\frac{\partial}{\partial y^0}K(k (x^0-y^0) ).\label{spin1result}
\end{align}
Thus, the dynamical response term in the limit $k\to 0$ satisfies
\begin{align}
	- \frac{\kappa}{15}a^2 \rho_b:- \frac{12\kappa}{15}a^2 \rho_b = 1:12.
\end{align}
This is consistent with $C_T$ for spin-$1$.
The dynamical response term can be recast as
\begin{align}
	T^{\rm dyn.}_{ij}(x^0,\mathbf{k}) &= -\frac{4}{3} \kappa a^2\rho_b h^{(s)}(x^0,\mathbf{k}) +8\kappa a^2\rho_b K(kx^0)h^{(s)}(0,\mathbf{k})
	 \notag 
	 \\
	 &+ 8\kappa a^2\rho_b \int^{x^0} dy^0K(k (x^0-y^0) ) \frac{\partial}{\partial y^0}h^{(s)}(y^0,\mathbf{k}),\label{dyn1}
\end{align}
and we can read off the fast response as
\begin{align}
	T^{\rm fast}_{ij}(x^0,\mathbf{k})  = -\frac{4}{3} \kappa a^2\rho_b h^{(s)}(x^0,\mathbf{k}).\label{qispin1}
\end{align}

\subsection{Contact term for the Maxwell field}

Here we consider the variation of \eqref{def:emstresstensor} with respect to the metric.
In the Coulomb gauge, $A_0=0$ and we fix $A_i$ and its conjugate momentum $\pi^i_A$ while varying $g^{\mu\nu}$.
The conjugate momentum is given by $\pi^i_A =F^{0i} $ in the TT gauge.
Hence, we set
\begin{align}
	\delta F^{i0} = \delta F_{ij} = 0.
\end{align}
Note that setting $\delta F_{0i} = 0$ gives an incorrect result, since the variation is defined in the canonical formalism.
This distinction is implicit for the conformal scalar and for Weyl fermions, since the scalar and spinor conjugate momenta do not carry spatial indices.
The second term in \eqref{def:emstresstensor} does not contribute to the TT component, so we only need to consider the variation of the first term.
The variation of the electric contribution yields
\begin{align}
	\int d^4y \delta g^{kl}(y)\frac{\delta}{\delta g^{kl}(y)} \langle  F_{i0}F_{j}{}^0 \rangle = \frac{2}{3}\kappa \langle E^2\rangle h_{ij},
\end{align}
and the magnetic counterpart is
\begin{align}
	\int d^4y \delta g^{kl}(y)\frac{\delta}{\delta g^{kl}(y)} \langle F_{im}F_{j}{}^m \rangle = -\frac{1}{3}\kappa h_{ij} \langle B^2\rangle.
\end{align}
Here the thermal average of the Maxwell field is given by
\begin{align}
	\langle F^{0i} F^{0j}\rangle &= \frac{1}{3}\delta^{ij}\langle E^2\rangle,
	\\
	\langle F_{ik} F_{jl}\rangle &= \frac{1}{3}(\delta_{ij}\delta_{kl} - \delta_{il}\delta_{kj})\langle B^2\rangle.
\end{align}
For photons we have $\rho = \langle E^2 \rangle =\langle B^2 \rangle$.
$E$ and $B$ appear to be asymmetric because we drop the second term in Eq.~\eqref{def:emstresstensor}. 
Thus the contact contribution of the Maxwell field is
\begin{align}
	\int d^4 x \left \langle \frac{\delta \hat T_{kl}}{\delta g^{ij}} \right \rangle_{\kappa =0} h^{ij} = P h_{kl}.
\end{align}
In summary, after the Weyl rescaling, we find
\begin{align}
	T^{\rm cont.}_{ij} = \kappa a^2 P h_{ij}.\label{contspint1}
\end{align}
Note that $P=2P_b$ with $P_b$ being the pressure of a relativistic bosonic degree of freedom.

\subsection{Summary of spin 1}

Combining Eqs.~\eqref{osterm}, \eqref{dyn1}, and \eqref{contspint1}, the shear tensor is found as
\begin{align}
	\pi_{ij} =  -24 a^2 \left(\frac{\kappa^2}{6}\rho_b\right) \int^{x^0} dy^0K(k (x^0-y^0) ) \frac{\partial}{\partial y^0}h_{ij}(y^0,\mathbf{k}) \label{eq:pi:sp:1}.
\end{align}
Using the Friedmann equation for a radiation-dominated universe with a single Maxwell field,
\begin{align}
	\frac{\kappa^2}{6}\rho_b = H^2.
\end{align}

\section{Consistency with previous one-loop analysis}

In earlier one-loop analyses~\cite{Ota:2023iyh,Frob:2025sfq,Ota:2024idm}, the thermal loop corrections to the gravitational-wave power spectrum were discussed and a secular term was reported. We must now address the consistency between that result and the present analysis. In these earlier works, both the graviton and the environmental radiation field were treated on the same footing as quantum fields. The calculation was intended to obtain the connected Keldysh Green function of primordial gravitational waves when they are coupled to a thermal environment.

Let us describe the same setup in the path integral formalism.
For a given action of the thermal radiation~$S_{\rm RD}[\chi;h]$ and the graviton in the TT gauge~$S_{g}[h]$, the total action $S[h,\chi;j] =S_{g}[h;j]+ S_{\rm RD}[\chi;h]$ couples to the graviton source $j$, and the path integral takes the form
\begin{align}
	e^{iW[j]} = \oint_{\hat d_h} \mathcal D h \oint_{\hat d_\chi} \mathcal D\chi e^{iS[h,\chi;j]},
\end{align}
where $\hat d_X$ is the density operator of $X$.
$\hat d_\chi$ is typically a thermal ensemble, while $\hat d_h$ is the adiabatic vacuum in the remote past of inflation.
$\oint$ denotes the closed time path integral with boundary conditions specified by $\hat d_X$.
We then computed the gravitational-wave power spectrum $P_h$, schematically written as
\begin{align}
	P_{h} = \frac{\delta^2 W[j]}{\delta j \delta j}.\label{72}
\end{align}
Note that the advanced index on $j$ is implicit.
In practice we neglected the vacuum gravitational propagators in comparison with the thermal ones, since the vacuum contribution is subdominant in the radiation background.
Two issues then arose:
\begin{itemize}
	\item secular growth,
	\item non-convergence of the thermal loop expansion.
\end{itemize}
The first issue arises from an inappropriate time boundary in the in-in calculation, which is resolved in the present work.
The non-convergence of the thermal loop expansion indicates the need for resummation; this is also achieved in this work as detailed below.

In this work, we split the full action into the gravity and radiation sectors and take only the partial trace over $\chi$:
\begin{align}
	e^{iW_{\rm th}[h]} =  \oint_{\hat d_\chi} \mathcal D\chi e^{iS_{\rm RD}[\chi;h]},
\end{align}
$h$ is treated as a nondynamical source for $\chi$, and $W_{\rm th}[h]$ is expanded in a series in $h$.
Up to quadratic order in $h$ our result is complete, because of the collisionless assumption for $\chi$ (i.e. $\chi$ is free).
Then the full path integral reduces to
\begin{align}
	e^{iW[j]} = \oint_{\hat d_h} \mathcal D h e^{iS_g[h;j]+iW_{\rm th}[h]}\,.\label{75}
\end{align}
Thus, $W_{\rm th}[h]$ is the generating functional of connected thermal Green functions; at the same time, it encodes the thermal vertex corrections for $h$.
The original problem of computing Eq.~\eqref{72} therefore reduces to a loop analysis of the thermal vertices with the dressed retarded and Keldysh propagators defined from Eq.~\eqref{75}, or, in the linear analysis discussed here, to an analysis of the dressed propagators themselves.

The similarity between the one-loop correction to $P_h$ and $W_{\rm th}$ can be seen as follows.
Writing $W_{\rm th}[h] = -h\Gamma h/2$, we have
\begin{align}
	S_g[h;j] + W_{\rm th}[h] = \frac{1}{2}h (G^{-1} - \Gamma) h + hj.
\end{align}
Then,
\begin{align}
	W[j] = \frac{1}{2}j(G^{-1} - \Gamma)^{-1}j =\frac{1}{2}j(G + G\Gamma G +\cdots)j.  
\end{align}
More precisely, we discussed the correction to $P_h = G^K$, which is expressed as $G^R\Gamma G^K$.
The vertex function that appears in one-loop $P_h$ can now be identified with $\Gamma$.
In fact, for each spin, we explicitly confirmed that the one-loop correction from the 4-point interaction for $P_h$ corresponds to the local response $T^{\rm os}_{\mu\nu} + T^{\rm cont.}_{\mu\nu}$ in the effective action.
In the previous analysis~\cite{Ota:2023iyh,Frob:2025sfq,Ota:2024idm}, the local Gibbs ensemble cancels the four-point interaction in the scalar case, while a recent work~\cite{Ota:2025rll} has shown that the ensemble is incompatible with small diffeomorphism invariance.
The IR one-loop spectrum in Refs.~\cite{Ota:2023iyh,Frob:2025sfq,Ota:2024idm} precisely corresponds to the infrared limit of $T^{\rm dyn.}_{\mu\nu}$.

\section{Conclusions}

This paper examines whether the spin-dependent coefficients of the two-point function, as shown in Eq.~\eqref{cft}, appear in the linear response of gravitational perturbations in thermal environments and modify the robust kinetic-theory results.
The short answer is both yes and no.
The coefficient $C_T$ in Eq.~\eqref{CTdef} is indeed encoded in the infrared mass-squared shift that arises in the dynamical response on the thermal time scale.
The mass-squared coefficients are found to be in the ratio $\rho_b:-3\rho_f:12\rho_b$ for a conformal scalar, a Weyl spinor, and a Maxwell field, respectively, where $\rho_b$ and $\rho_f$ are the energy densities of a relativistic bosonic and fermionic degree of freedom.
This is analogous to the zero-temperature case, $1:3:12$ \cite{OsbornCFTNotes}.
The minus sign for the Weyl fermion is from the sign of the finite temperature fermion Keldysh propagator.
However, this IR feature of the dynamical response is completely confined within the real-time response, it is canceled by the local responses coming from the on-shell projection in the Einstein equation and from the explicit source dependence of the stress tensor.
We have demonstrated this cancellation explicitly for each spin.
For a general mixture of spins, we then obtain the universal result
\begin{align}
	h_{ij}'' + 2 \mathcal H h_{ij}' + k^2 h_{ij}   =-24 \mathcal H^2 \int^{x^0} dy^0K(k (x^0-y^0) ) \frac{\partial}{\partial y^0}h_{ij}(y^0,\mathbf{k}).\label{effectiveEOM}
\end{align}
Weinberg's classical kinetic-theory result is thus reproduced.

This study provides, to our knowledge, the first consistent calculation of the gravitational hard thermal loop linear response in the real-time formalism, with the correct time boundary and without any \textit{ad hoc} prescription for the local contact contribution~(see \cite{Caron-Huot:2007cma} for the real-time approach in gauge theory HTL analysis).
From the symmetry point of view, the cancellation of the fast response is in some sense a guaranteed outcome, since the kinetic-theory result is known to be consistent with the Ward identities of small diffeomorphisms~\footnote{It is conventional wisdom that the graviton is massless because of diffeomorphism invariance.
This is straightforward in the zero temperature limit, but the presence of a thermal bath makes the statement more subtle.
In fact, early references claimed that the same conclusion holds in the thermal case based on so-called Jeans swindle.
As a rigorous statement, the importance of the off-shell identity was emphasized in Ref.~\cite{Ota:2025rll}, where it was argued that a genuine symmetry test is only available for the off-shell stress tensor, and the effective mass of the graviton is zero for the on-shell dynamics.
}.
However, the symmetry argument alone does not fix the overall normalization, which we have verified explicitly in this work.
Although reproducing an established result might appear conservative, the exact cancellation within quantum field theory is highly nontrivial.
In particular, the mass shift of order $\mathcal O(T^4)$ corresponds to the power-law divergence part at zero temperature.
Although such a power-law divergence can be removed by renormalizing physical constants at zero temperature, the finite-temperature contribution is itself finite.
Geometric counterterms in the Einstein-Hilbert action cannot generate a $T^4$ mass shift, even after matching to the background Friedmann equation; the cancellation must therefore be realized within linear response theory.
In our analysis for fields of various spin, a non-vanishing on-shell projection $T^{\rm os}_{\mu\nu}$ is essential to demonstrate the exact cancellation.
This implies that a dynamical background is required to ensure the universal cancellation.

The reproduction of the kinetic-theory result in the HTL limit provides a robust foundation for the study of generic nonthermal environments that cannot be described by the HTL approximation alone.
In this direction, our real-time approach has a clear advantage over the traditional imaginary-time formalism, which is restricted to strictly thermal systems.
Our result can be used as a benchmark for extending the analysis to nonthermal or nontrivial zero-temperature dynamics during cosmic inflation or the preheating and reheating phases.
In particular, our linear response analysis with the effective action will be useful for discussing loop effects from the ultra-slow-roll inflation relevant to primordial black holes and induced gravitational waves. Our formulation indicates that the appropriate choice of time boundary for the retarded stress-tensor propagator is not necessarily compatible with that for the primary fields. This tension may be a generic feature of infrared calculations in the in-in formalism.
Promising future directions include extending our framework to scalar-vector-tensor perturbations.
A genuine one-loop analysis with the dressed Green functions and with thermal non-Gaussianities of the thermal vertices would also be an interesting application.

\acknowledgments
AO was supported in part by the National Natural Science Foundation of China under Grant No. 12547101 and 12347101 and 12403001, and New Chongqing YC Project CSTB2024YCJH-KYXM0083. H.Y.Z. and Y.Z is supported by IBS under the project code, IBS-R018-D3. 
AO is grateful to Yingying Lan for her generous support.

\appendix

\section{Covariant kinetic theory in the interaction picture}\label{sec:Kinth}

The main text analyzes the gravitational-wave response in quantum field theory in the real-time formalism.
This appendix provides details of the classical kinetic counterpart.
We review Ref.~\cite{Francisco:2016rtf} and derive the linear response in kinetic theory, while also elaborating on several technical aspects that were not discussed there in detail.

\subsection{Vlasov equation}

In kinetic theory the phase-space distribution function $f$ determines the stress tensor:
\begin{align}
	T_{\mu\nu} = 2 \int \frac{d^4 p}{(2\pi)^4\sqrt{-g}} (2\pi)\delta(g^{\alpha \beta} p_{\alpha}p_{\beta}+m^2)\theta(E) p_{\mu} p_{\nu} f(x^{\gamma},p_{\delta}),\label{Tmunuinkinetic}
\end{align}
where $x$ and $p$ denote the position and its conjugate momentum, the delta function imposes the mass shell, and the step function selects the positive-energy branch, $E=-p_{0}\geq 0$.  
The weak energy condition requires $T_{00}\geq 0$, and Eq.~\eqref{Tmunuinkinetic} ensures that this component is positive-definite, independently of the metric signature.
$f$ is understood as the sum over the relativistic degrees of freedom:
\begin{align}
	f = \sum_{i} f_i = g_* f_0,
\end{align}
with the distribution function for the spin-0 particle $f_0$.
Then, computing Eq.~\eqref{Tmunuinkinetic} in kinetic theory is an alternative way to evaluate the effective stress tensor~\eqref{deflinres}.

\medskip
The function $f$ may be defined off shell in the full eight-dimensional phase space.  
It is convenient to introduce the covariant on-shell distribution
\begin{align}
	F(x^{\gamma},p_{\delta}) \equiv w(x^{\gamma},p_{\delta}) f(x^{\gamma},p_{\delta}),\label{defonoff}
\end{align}
with the window function
\begin{align}
	w(x^{\gamma},p_{\delta}) \equiv 2(2\pi)\delta(g^{\alpha \beta} p_{\alpha}p_{\beta}+m^2)\theta(E).\label{onshellcond}
\end{align}
In general $w$ depends explicitly on $x$ through the metric.

\medskip
The Hamiltonian for a point particle in the eight-dimensional canonical phase space is
\begin{align}
	\mathbb H(x^\mu,p_\nu) = \frac{1}{2} g^{\mu \nu} p_\mu p_\nu ,\label{polyakov}
\end{align}
with Hamilton equations
\begin{align}
	\frac{\partial \mathbb H}{\partial x^\mu} &=- \frac{d p_\mu}{d\lambda} =  \frac{1}{2}p_\alpha p_\beta \partial_\mu g^{\alpha \beta}, \label{solgeod}
	\\
	\frac{\partial \mathbb H}{\partial p_\mu} &=\frac{d x^\mu}{d\lambda}=   g^{\mu\nu}p_\mu.
\end{align}
Along a geodesic $\mathbb H$ is conserved, so the motion remains on a given mass-shell, although this shell need not correspond to the physical mass of a real particle, as in Eq.~\eqref{onshellcond}.

\medskip
Defining the covariant number flux
\begin{align}
	N^\mu \equiv \int \frac{d^4 p}{(2\pi)^4\sqrt{-g}}p^\mu F,\label{Nmuinkinetic}
\end{align}
one finds
\begin{align}
	\nabla_\mu N^\mu &= \frac{1}{\sqrt{-g}}\partial_\mu \left( \sqrt{-g}\int \frac{d^4 p}{(2\pi)^4\sqrt{-g}}p^\mu F\right)
	\notag 
	\\
	& =\int \frac{d^4p}{(2\pi)^4\sqrt{-g}}\left[ p^\mu \partial_\mu  F+  p_\rho (\partial_\mu g^{\mu \rho} )    F\right]
		\notag 
	\\
	& =\int \frac{d^4p}{(2\pi)^4\sqrt{-g}}\left[\frac{dx^\mu}{d\lambda}\frac{\partial F}{\partial x^\mu } + \frac{dp_\mu}{d\lambda}\frac{\partial F}{\partial p_\mu } \right]
	\notag 
	\\
	& =   \int \frac{d^4p}{(2\pi)^4\sqrt{-g}}\frac{dF}{d\lambda}.
\end{align}
Therefore, the conservation law $\nabla_\mu N^\mu=0$ applied to an arbitrary covariant volume in momentum space follows from the Vlasov equation
\begin{align}
\frac{dF}{d\lambda} = 0. \label{eq:vlasov}
\end{align}
Because the on-shell condition is preserved along $\lambda$, i.e.\ $d\mathbb H/d\lambda=0$, one may write
\begin{align}
\frac{dF}{d\lambda} = w \frac{df}{d\lambda}. \label{eq:vlasov2}
\end{align}
Hence
\begin{align}
	\int \frac{d^4p}{(2\pi)^4\sqrt{-g}}\frac{dF}{d\lambda} = \int \frac{d^3p}{(2\pi)^3\sqrt{-g}p^0} \frac{df}{d\lambda} ,\label{eq11}
\end{align}
which might suggest that $df/d\lambda=0$ on shell.  
Since the $\lambda$-derivative is tangent to a given mass-shell~\cite{Acuna-Cardenas:2021nkj}, we expect
\begin{align}
	\left .\frac{df}{d\lambda} \right |_{\rm on-shell} = \frac{d}{d\lambda} \left. f\right|_{\rm on-shell},\label{onshell}
\end{align}
the quantity that appears in standard cosmological kinetic theory.

\subsection{Stress tensor in kinetic theory}\label{sec:htl}

As a preliminary step, and in order to connect with the standard HTL framework, we first solve the linear response problem in flat spacetime.  
Because the Vlasov equation for massless particles is invariant under Weyl rescaling, the result can subsequently be promoted to the FLRW background.

\medskip
In linear response, we perturb $F$ by the metric fluctuation and reconstruct the stress tensor~\eqref{Tmunuinkinetic}.  
The deviation of $F$ obeys the Vlasov equation
\begin{align}
	\frac{dp_\mu}{d\lambda}\frac{\partial F}{\partial p_\mu } + \frac{dx^\mu}{d\lambda}\frac{\partial F}{\partial x^\mu } = 0.
\end{align}
We begin with a flat background and write
\begin{align}
	g^{\mu\nu} = \eta^{\mu\nu} + \kappa h^{\mu\nu},\label{flatmetricdef}
\end{align}
so that
\begin{align}
	\frac{dx^\mu}{d\lambda}\frac{\partial F}{\partial x^\mu } =  g^{\mu\nu}p_{\nu} \partial_\mu F = (\eta^{\mu\nu}p_\nu + \kappa h^{\mu\nu}p_\nu)\partial_\mu F.
\end{align}
Using Eq.~\eqref{solgeod} for the momentum term, the perturbed Vlasov equation becomes
\begin{align}
		\hat M F &= \kappa \hat L F,\label{pertvlasov}
\end{align}
where we define
\begin{align}
	\hat M &\equiv p \cdot \partial =\eta^{\mu\nu}p_\nu \partial_\mu,
	\\
	\hat L &\equiv \frac{1}{2} \left(  p_\mu p_{\nu} \partial_\lambda h^{\mu \nu }  \frac{\partial }{\partial p_\lambda } - h^{\mu\nu}(p_\nu \partial_\mu + p_\mu \partial_\nu) \right).
\end{align}
So far the Vlasov equation is exact in the expansion in $\kappa$.
We solve this equation order by order in $\kappa$. At leading order,
\begin{align}
\hat M F^{(0)} = 0. \label{zerothvlasov}
\end{align}
If $F^{(0)}$ is taken to be independent of $x^\mu$ as the background is flat, it remains constant along free trajectories.  
Equation~\eqref{pertvlasov} can then be integrated formally to give
\begin{align}
	F = \sum_{n=0}^{\infty} \left( \kappa \hat M^{-1} \hat L \right)^n F^{(0)} = \frac{1}{1 - \kappa \hat M^{-1}\hat L}F^{(0)}.
\end{align}
Once $F^{(0)}$ is fixed at zeroth order in the metric perturbation, all higher-order corrections follow automatically.  
The stress tensor is then given by
\begin{align}
		T_{\mu\nu} &= \int \frac{d^4 p}{(2\pi)^4\sqrt{-g}} p_\mu p_\nu  \frac{1}{1 - \kappa \hat M^{-1}\hat L}F^{(0)}
		\notag 
		\\
		&=    \int \frac{d^4 p}{(2\pi)^4\sqrt{-g}} F^{(0)}\left[ p_\mu p_\nu +  \frac{\kappa}{2} \left( 
  \frac{p_\mu p_\nu p_\rho p_\sigma}{(p \cdot \partial)^2}  \partial \cdot \partial
 -  \frac{ \partial_{\langle \mu} p_\nu p_\rho p_{\sigma\rangle }}{p \cdot \partial} 
   \right) h^{\rho\sigma} \right] + \mathcal O(\kappa^2)
   		\notag 
		\\
		&=    \rho \int \frac{d^2 Q}{4\pi} \left[\left(1 +\frac{\kappa}{2}\eta_{\rho \sigma}h^{\rho \sigma}\right) Q_\mu Q_\nu +  \frac{\kappa}{2} \left( 
  \frac{Q_\mu Q_\nu Q_\rho Q_\sigma}{(Q \cdot \partial)^2}  \partial \cdot \partial
 -  \frac{ \partial_{\langle \mu} Q_\nu Q_\rho Q_{\sigma\rangle }}{Q \cdot \partial} 
   \right) h^{\rho\sigma} \right] + \mathcal O(\kappa^2),\label{kernelexpand}
\end{align}
where we defined the null 1-form $Q_\mu = -p_\mu/p_0$, the cyclic permutation
\begin{align}
	\partial_{\langle \rho} p_{\sigma} p_\mu p_{\nu \rangle} = \partial_{\rho} p_{\sigma} p_\mu p_{\nu} + \partial_{\sigma}  p_\mu p_{\nu} p_{\rho} + \partial_{\mu}  p_{\nu} p_{\rho} p_\sigma + \partial_{\nu} p_{\rho}  p_\sigma p_{\mu},
\end{align}
and the energy density
\begin{align}
	\rho = - \int \frac{d^4p}{(2\pi)^4}p^0p_0 F^{(0)}= - \int \frac{d^3p}{(2\pi)^3}p_0  f^{(0)}.\label{defrhoflrw}
\end{align}
Eq.~\eqref{kernelexpand} is the classical counterpart of the effective stress tensor discussed in the main text.

\subsection{Weyl invariance of the Vlasov equation}

The perturbed FLRW metric is given by the Weyl rescaling of Eq.~\eqref{flatmetricdef}:
\begin{align}
	g^{\mu\nu} = a^{-2}(\eta^{\mu\nu} + \kappa h^{\mu\nu}).
\end{align}
In the flat FLRW background, a straightforward calculation yields 
\begin{align}
	\frac{dp_\mu}{d\lambda} \frac{\partial F}{\partial p_\mu}
	&= \frac{1}{a^2 }   p_\alpha p_{\beta}  \left ( a^2 \mathcal H \delta_\mu^0 g^{\alpha \beta}  - \frac{1}{2} \kappa \partial_\mu  h^{\alpha \beta} \right) \frac{\partial F}{\partial p_\mu},\label{newres1}
\end{align}
and 
\begin{align}
	\frac{dx^\mu}{d\lambda}\frac{\partial F}{\partial x^\mu} 
	&=a^{-2} (\eta^{\mu\nu}p_\nu + \kappa  h^{\mu\nu} p_\nu)\partial_\mu F.
	\label{newres2}
\end{align}
Since $F$ is a distribution with off-shell support, we should not immediately set 
\begin{align}
	   p_\alpha p_{\beta} g^{\alpha\beta}  \mathcal H  \frac{\partial F}{\partial p_0} = 0.
\end{align}
In fact, with a test function $X$, 
\begin{align}
	  X  p_\alpha p_{\beta} g^{\alpha\beta}  \mathcal H  \frac{\partial F}{\partial p_0} = - 2 X p_{\beta} g^{0\beta}  \mathcal H  F \neq 0.\label{van}
\end{align}
The mass-shell scales as $a^{-2}$ in the delta function, so $F$ scales as $F \to a^2 F$.  
Including this scaling factor, Eq.~\eqref{newres2} is replaced by
\begin{align}
	 (\eta^{\mu\nu}p_\nu + \kappa  h^{\mu\nu} p_\nu)\partial_\mu F + 2a^2\mathcal H  g^{0\nu}p_\nu F.
\end{align}
The second term cancels Eq.~\eqref{van}, showing that the Vlasov equation for $F$ is Weyl invariant.

\section{Kernel structure and homogeneous solutions}\label{appkernel}

In this appendix, we provide details of the kernel structure of Eq.~\eqref{kernelexpand}, and explain how the homogeneous solution~\eqref{bcextra} arises.
From here on we focus on gravitational waves: $h^{\mu0}=\delta_{ij}h^{ij} = \partial_i h^{ij}=0$.
The relevant spatial components of Eq.~\eqref{kernelexpand} in Fourier space are given by
\begin{align}
	T_{ij}(\tau, \mathbf{k}) &= \frac{\kappa}{2} \rho \int \frac{d^2 Q}{4\pi} \left( \left(\partial_0^2 + k^2 \right) \frac{Q_i Q_j Q_k Q_l}{\left(\partial_0 + i \vec Q\cdot \mathbf{k}\right)^2} \right)h^{kl}(\tau,\mathbf{k}), \label{Pijcovariant}
\end{align}
with $k\equiv |\mathbf{k}|$.  
Its TT projection is 
\begin{align}
	 - \frac{\kappa}{2} \rho \int_{-1}^{1} \frac{d\mu}{2} \frac{\partial_0^2 + k^2}{(\partial_0 + i k \mu)^2} \frac{(1 - \mu^2)^2}{4}h_{ij}(\tau,\mathbf{k}). \label{412}
\end{align}
This second-order operator can be rearranged in terms of first-order operators:
\begin{align}
	-\frac{3}{4}  \int_{-1}^{1} \frac{d\mu}{2}
	(1 - \mu^2)^2\frac{ \partial_0^2+ k^2}{(\partial_0 + i k \mu)^2} = -2 +3 \int_{-1}^{1} \frac{d\mu}{2}
	(1 - \mu^2)^2 \frac{\partial_0 }{\partial_0 + i k \mu}  \label{result:self}.
\end{align}
Finding the kernel function is therefore equivalent to solving
\begin{align}
	\partial_0 h = (\partial_0 + ik\mu)s.
\end{align}
The solution of this differential equation is the sum of homogeneous and particular parts:
\begin{align*}
	s = e^{-ik\mu\tau} \sigma_{ij} + \int^{\tau}_{\tau_0} d\bar \tau \, e^{ik\mu (\bar \tau -\tau ) } \partial_{\bar \tau} h_{ij}(\bar \tau),
\end{align*}
where $\sigma_{ij}$ is an integration constant that is TT.  
Substituting this integration constant in Eq.~\eqref{412} and integrating over $\mu$, we obtain Eq.~\eqref{bcextra}.
Thus, in the kinetic-theory approach, Eq.~\eqref{kernelexpand} is left unintegrated in time, so the time boundary is not specified.
One may therefore fix $\sigma$ so as to respect the infrared symmetry in the equation of motion.

\bibliography{biblio.bib}{}
\bibliographystyle{unsrturl}

\end{document}